\documentclass[amsmath,amssymb,aps,prc,twocolumn,superscriptaddress,showpacs,preprintnumbers]{revtex4}
\usepackage{graphicx}
\usepackage{multirow}
\usepackage{times}
\usepackage{color}
\usepackage[normalem]{ulem}  % \sout{old text} for strikeout

\renewcommand\sout{\bgroup \color{red} \ULdepth=-.5ex \ULset}
\DeclareMathOperator{\tr}{tr}
\DeclareMathOperator{\diag}{diag}
\DeclareMathOperator{\cn}{cn}
\DeclareMathOperator{\Ai}{Ai}

\newcommand{\half}{\frac{1}{2}}
\newcommand{\tilA}{\tilde{A}}
\newcommand{\tilp}{\tilde{p}}
\newcommand{\del}{\partial}
\newcommand{\deltau}{\partial_\tau}
\newcommand{\deltheta}{\partial_\theta}
\newcommand{\tiltheta}{\tilde{\theta}}
\newcommand{\calF}{\mathcal{F}}
\newcommand{\calA}{\mathcal{A}}
\newcommand{\calB}{\mathcal{B}}
\newcommand{\calD}{\mathcal{D}}
\newcommand{\calE}{\mathcal{E}}
\newcommand{\calO}{\mathcal{O}}
\newcommand{\calL}{\mathcal{L}}

\newcommand{\conj}{\text{c.c.}}

\newcommand{\rp}{\right)}%right parentheses
\newcommand{\lp}{\left(}%left parentheses
\newcommand{\rbb}{\right]}%right box bracket 
\newcommand{\lbb}{\left[}%left box bracket
\newcommand{\rtB}{{B_\text{eff}^{1/2}}}
\newcommand{\PSfig}[2]{\includegraphics[width=#1]{#2}}
\begin{document}
% number in the upper right hand corner of the title page in preprint mode.
\preprint{KUNS-2598, YITP-15-101}

\title{
Parametric Instability of Classical Yang-Mills Fields in an Expanding Geometry
}

\author{Shoichiro Tsutsui}
\email[]{tsutsui@ruby.scphys.kyoto-u.ac.jp}
%\thanks{}
\affiliation{Department of Physics, Faculty of Science, Kyoto University,
Kyoto 606-8502, Japan}
%%%%%
%%%%%
\author{Teiji Kunihiro}
\affiliation{Department of Physics, Faculty of Science, Kyoto University,
Kyoto 606-8502, Japan}
%%%%%
\author{Akira Ohnishi}
\affiliation{Yukawa Institute for Theoretical Physics, Kyoto University,
Kyoto 606-8502, Japan}

\date{\today}
\pacs{03.50.-z, 11.15.Kc, 12.38.Mh}

\begin{abstract}
We investigate the instability of classical Yang-Mills field in an expanding geometry under a color magnetic background field within the linear regime.
We consider homogeneous, boost-invariant and time-dependent color magnetic
fields simulating the glasma configuration. 
We introduce the conformal coordinates which enable us to map an expanding problem approximately
into a nonexpanding problem.
We find that the fluctuations with finite longitudinal 
momenta can grow exponentially due to parametric instability.
Fluctuations with finite transverse momenta
can also show parametric instability,
but their momenta are restricted to be small.
The most unstable modes start to grow exponentially in the early stage of the dynamics and they may affect the thermalization in heavy-ion collisions.
\end{abstract}

\maketitle

%%%%%%%%%%%%%%%%%%%%%%%%%%%%%%%%%%%%%%%%%%%%%%%%%%%%%%%%%%%%%%
\section{Introduction}

%\textbf{GENERAL BACKGROUND:}
Recent developments in physics of high-energy heavy-ion collisions have unveiled the property of the quark-gluon plasma (QGP) and have raised a puzzle in the pre-equilibrium dynamics before the QGP formation.
High-energy heavy-ion collision experiments have been performed extensively at Relativistic Heavy-Ion Collider (RHIC) at Brookhaven National Laboratory and Large Hadron Collider (LHC) at CERN. 
Theoretical analyses of data based on hydrodynamical approaches are found to be successful in explaining various observables such as hadron momentum spectra and collective flows in the nucleus nucleus collisions~\cite{Heinz05,Teaney09,Hirano12}.
The hydrodynamical analyses suggest that QGP formed at RHIC and LHC would be a strongly interacting fluid rather than a weakly interacting gas.
However, it should be noted that the hydrodynamical models still pose a puzzle.
The early thermalization required by the hydrodynamical analyses appears inconsistent with perturbative QCD results,
such as those given by the bottom-up scenario~\cite{Baier01,Baier02}.
To resolve the puzzle, 
we need deeper understanding of the dynamical nature of the initial stage of the created matter, say, in view of far-from-equilibrium dynamics of non-Abelian gauge theory.

%\textbf{INSTABILITIES IN HIC:}
The initial stage of relativistic heavy-ion collisions may be well described by the effective theory based on the notion of the color glass condensate~\cite{McLerran:1993ni,Gelis10a}.
It is shown that the longitudinal color flux tubes which consist of both color electric 
and magnetic fields should be formed right after the collision of two nuclei.
This state with the longitudinal color-flux tube is called glasma~\cite{Lappi06}.
The time evolution of the low-energy sector is well described by classical Yang-Mills (CYM) theory as the first approximation although quantum fluctuations can 
significantly affect the dynamics.
For instance, quantum fluctuations would trigger plasma instabilities,
which in turn might cause the emergence of chaoticity, turbulent spectra, rapid particle production and 
thermalization~\cite{Berges09,Kunihiro08,Kunihiro10,Iida13}.

Glasma instabilities in heavy-ion collisions have been extensively discussed.
Among them, longitudinal fluctuations are found to cause an instability in glasma
~\cite{Romatschke06a,Romatschke06b,Berges08,Epelbaum14,Rebhan:2008uj,Attems:2012js}.
One possible underlying mechanism of the instabilities 
is non-Abelian analog of the Weibel instability~\cite{Weibel59,Mrowczynski88,Arnold03,Arnold05}, 
which leads to an exponential amplification of the color magnetic field and current in anisotropic systems.
The longitudinal color magnetic field may also induce
the Nielsen-Olesen instability~\cite{Nielsen78,Fujii08,Fujii09,Iwazaki09,Kurkela:2011ti,Kurkela:2011ub},
which is the exponential growth of gauge fluctuations caused by the anomalous Zeeman effect in spin-1 systems.

%\textbf{RELATED STUDIES:}
Recently, it has been suggested that
the CYM field under a time-dependent homogeneous color magnetic 
background field shows instability 
in a nonexpanding geometry~\cite{Berges12a}.
It has been clarified that the origin of the instability is \textit{parametric instability} and
the CYM field in a homogeneous background has multiple instability bands extending to the transverse as well as the longitudinal momentum region~\cite{Tsutsui:2014rqa}.
The parametric instability is ubiquitous in physics from classical mechanical problems on a pendulum 
to quantum field theories~\cite{Berges02c},
as well as in the cosmic inflation~\cite{Amin:2014eta}.
These results suggest that the parametric instability may also play an essential role in the thermalization process in heavy-ion collisions.

From a phenomenological point of view,
it is natural to ask ourselves whether and how the parametric instability
found in a nonexpanding geometry persists in an expanding geometry.
In general, background fields decrease in time due to the expansion, 
so one may expect that the instability would disappear or at least become less significant than in the nonexpanding geometry.
However, the detailed studies of a \textit{scalar field theory} suggest 
that the parametric instability can keep significance even in an expanding geometry~\cite{Berges12b,Dusling:2012ig}.

%\textbf{THIS ARTICLE:}
In this article,
we investigate the instability of the CYM field in an expanding geometry in the linear regime.
Specifically, we focus on the system with the longitudinal color magnetic field which is homogeneous, boost-invariant and time-dependent.
We also assume the one-dimensional Bjorken expansion.
The background field considered here damps due to the expansion, and at the same time, oscillates in time.
We introduce a natural chronological variable called conformal time 
which enables us to map an expanding problem into a nonexpanding problem.
Then the relevant equation is found to have a form of a temporally periodic-driven 
equation like, $\ddot{f}+p^2 f+\lambda\cn^2(t;k)f=0$.
This equation involves the Jacobi elliptic function $\cn(t;k)$, 
and is called Lam\'{e}'s equation for a constant momentum $p$ and a constant coefficient $\lambda$. Lam\'{e}'s equation shows an exponential instability at $\lambda=-1$ and $2$ for a small momentum $p$.
In the expanding case, the equation of motion is found to contain an effective momentum, which
is a function of the initial momentum and the conformal time and has decreasing longitudinal and increasing transverse components in time.
As a result, fluctuations with finite longitudinal momenta tend to be unstable due to the parametric instability,
while those with a finite transverse momentum can also show instability but their momenta are restricted to small values.

%CONSTRUCTION
%\textbf{Paper organization:}
This article is organized as follows.
In Sec.~\ref{sec:EOM}, we introduce the conformal time and
derive equations of motion of CYM field in an expanding geometry.
We also discuss properties of a homogeneous, boost-invariant and time-dependent background field.
In Sec.~\ref{sec:Floquet}, we give a brief outline of the Floquet theory 
and the instability band structure of Lam\'{e}'s equation as a useful tool to understand the expanding problem.
In Sec.~\ref{sec:Stability}, we show the results of the linear stability analysis.
Finally, we summarize our results and give concluding remarks in Sec~.\ref{sec:Conc}.
In Appendix A, we show the explicit form of the linearized equation of motion of fluctuations.
In Appendix B, we comment on the origin of a linearly divergent solution. In appendix C, we give a quantitative discussion on the growth rate.

%%%%%%%%%%%%%%%%%%%%%%%%%%%%%%%%%%%%%%%%%%%%%%%%%%%%%%%%%%%%%%
%%%%%%%%%%%%%%%%%%%%%%%%%%%%%%%%%%%%%%%%%%%%%%%%%%%%%%%%%%%%%%
\section{Classical Yang-Mills field in an expanding geometry}\label{sec:EOM}
In this section, we derive the equation of motion (EOM) of the CYM field in an expanding geometry under a color magnetic background field, which is homogeneous, boost invariant, and time-dependent.
In Sec.~\ref{subsec:setup}, we introduce the conformal time for the sake of mapping the expanding problem into a nonexpanding problem.
In Sec.~\ref{subsec:bg}, we discuss properties of the background field.
In Sec.~\ref{subsec:fluc}, we show an explicit form of the linearized EOM of fluctuations.

%%%%%%%%%%%%%%%%%%%%%%%%%%%%%%%%%%%%%%%%%%%%%%%%%%%%%%%%%%%%%%
\subsection{Expanding geometry and conformal time}\label{subsec:setup}
We introduce the proper time $\tau$ and the space-time rapidity $\eta$ to describe boost-invariant longitudinal expansion.
They are defined by
\begin{align} 
	\tau &= \sqrt{t^2 - z^2}, \\
	\eta &= \half \log\frac{t+z}{t-z}.
\end{align} 
The action of the CYM field in the $\tau$-$\eta$ coordinate is given by 
\begin{align}
	S 
	&= 
	\int d\tau d^2x_\perp d\eta \sqrt{-g} 
	\lp
	-\frac{1}{4}g^{\mu\nu}g^{\lambda\sigma} \calF_{\mu\lambda}^a \calF_{\nu\sigma}^a
	\rp.
	\label{action}
\end{align}	
Here, $\calF_{\mu\nu}^a = \del_\mu \calA^a_\nu + \del_\nu \calA^a_\mu + f^{abc}\calA^b_\mu \calA^c_\nu$
is the field strength tensor and $f^{abc}$ is the structure constant.
The coupling constant is included in the definition of the gauge field $\calA^a_\mu$. 
%\sout{$g^{\mu\nu}$ is the metric and its}
The explicit form of the metric $g^{\mu\nu}$ is given by
\begin{align}
	g^{\mu\nu} &= \diag(1,-1,-1,-1/\tau^2), \\
	g_{\mu\nu} &= \diag(1,-1,-1,-\tau^2), \\
	\det{g_{\mu\nu}} &\equiv g = -\tau^2. 
\end{align}	
Gauss's law in the $\tau$-$\eta$ coordinate is expressed as
\begin{align}
	\calD_i \calE^a_i + \calD_\eta \calE^a_\eta = 0,
	\label{GL}
\end{align}	
where the color electric fields are defined as
$\calE_i = \tau \deltau A_i (i=x,y)$ and
$\calE_\eta = \frac{1}{\tau} \deltau A_\eta$. 
The covariant derivative is given by $\calD_\mu = \del_\mu - i \calA_\mu$.

It is useful to see how the gauge field decreases by the expansion of the system.
The Bjorken's solution of the longitudinally
expanding hydrodynamics  tells us that the energy density decreases as $\epsilon \propto \tau^{-4/3}$~\cite{Bjorken:1982qr}.
Provided that this behavior also applies to the gauge field,
we expect that the amplitude of the gauge field decreases as 
$\calA \propto \tau^{-1/3}$, since the energy density of CYM fields
contains terms proportional to $\calA^4$. 
Motivated by this observation, let us consider the following time-dependent scale transformation~\cite{Berges12b}:
\begin{align}
	\deltau &= \tau^{-1/3}\del_\theta , \label{trans_del} \\
	\calA^a_i &= \tau^{-1/3}A^a_i ,\label{trans_Ai}\\
	\calA^a_\eta &= \tau^{-1/3}A^a_\eta ,\label{trans_Aeta}
\end{align}
where $\calA$ and $A$ denote the gauge fields in the $\tau$-$\eta$
coordinate and in the scaled coordinate, respectively.
We introduce the new chronological variable called conformal time $\theta$.
The relation between the conformal time and the proper time is explicitly given by 
%%
%%
%\begin{align}
%	\tau(\theta) = \lp \frac{2}{3}\lp \theta - \theta_0 \rp + \tau_0^{2/3} \rp^{3/2},
%\end{align}
%
%
%or choosing $\tau_0^{2/3}=2\theta_0/3$, we get
%
%
\begin{align}
\tau(\theta) = \lp \frac{2}{3} \theta \rp^{3/2} \equiv \tiltheta^{3/2} .
 \label{confornal_time}
\end{align}
Taking the Schwinger gauge $\calA^a_\tau = 0$,
the action in the new coordinate reads
\begin{align}
	S 
	&= 
	\int d\theta d^2x_\perp d\eta \Bigg[
	\half  \lp\deltheta A^a_i\rp^2
	-\half \frac{1}{9\tiltheta^2} A^a_i A^a_i 
	-\frac{1}{4}\tilde{F}^a_{ij}\tilde{F}^a_{ij} \notag\\
	&+\frac{1}{\tiltheta^3} \lp \half \lp\deltheta A^a_\eta\rp^2 
	-\half  \frac{7}{9\tiltheta^2} A^a_\eta A^a_\eta
	-\half \tilde{F}^a_{\eta i}\tilde{F}^a_{\eta i}
	\rp 
	\Bigg] .
	\label{action_conformal}
\end{align}
From now on, the capital letters denote all space components as $I,J,\dots = x,y,\eta$ and the lower letters
%\sout{denote} 
only transverse components as $i,j,\dots = x,y$.
The field strength tensor and covariant derivative in the new coordinate 
are defined by 
\begin{align}
\tilde{F}_{\mu\nu}^a 
&= \tilde{\del}_\mu A^a_\nu + \tilde{\del}_\nu A^a_\mu + f^{abc}A^b_\mu A^c_\nu \ ,\\
\tilde{D}_I^{ab} &= \tilde{\del}_I\delta^{ab} + f^{acb} A^c_I \ ,\\
\tilde{\del}_I &=\tau^{1/3}\del_I=\tiltheta^{1/2}\del_I \ .
\end{align}

In the action Eq.~\eqref{action_conformal},
the overall time dependence coming from the metric is absorbed into the 
%\sout{time dependent mass}\com{the coefficient of the quadratic terms}
the coefficient of the quadratic terms
and the spatial derivatives, 
%\com{so that}
so that the action 
%\sout{gets to have}\com{has}
has a similar form to that in a nonexpanding 
%\sout{space-time}\com{geometry}.
geometry.
The Euler-Lagrange equation is given by
\begin{align}
	\deltheta^2 A^a_i + \frac{1}{9\tiltheta^2}A^a_i - \frac{1}{\tiltheta^3} \tilde{D}_\eta \tilde{F}_{\eta i}^a - \tilde{D}_j \tilde{F}_{ji}^a &= 0,
	\label{eom_i} \\ 
	\deltheta \lp \frac{1}{\tiltheta^3} \deltheta A^a_\eta \rp + \frac{7}{9\tiltheta^5}A^a_\eta - \frac{1}{\tiltheta^3} \tilde{D}_i \tilde{F}_{i\eta}^a &= 0.
	\label{eom_eta}
\end{align}
Substituting Eqs.~\eqref{trans_del}, \eqref{trans_Ai} and \eqref{trans_Aeta} into Eq.~\eqref{GL}, 
we see that Gauss's law is expressed in terms of the conformal variables as
\begin{align}
	\tilde{D}_i \lp \deltheta A^a_i - \frac{A^a_i}{2\theta}  \rp 
	+ \frac{1}{\tiltheta^3} \tilde{D}_\eta \lp \deltheta A^a_\eta - \frac{A^a_\eta}{2\theta}  \rp
	=0	.
	\label{GL_conformal}
\end{align}
%
%

%%%%%%%%%%%%%%%%%%%%%%%%%%%%%%%%%%%%%%%%%%%%%%%%%%%%%%%%%%%%%%
\subsection{Background field configuration}\label{subsec:bg}
In this subsection, 
we introduce the gauge configuration which makes a homogeneous and boost-invariant longitudinal color magnetic field.
In order to extract the essential ingredients of the non-Abelian gauge theory
in a simple manner,
we consider color SU(2).
%\comm{
The color magnetic field in the $\tau$-$\eta$ coordinate is defined by
\begin{align}
	\calB_I^a = \epsilon_{IJK} \lp \del_J \calA_K^a  -\half \epsilon^{abc} \calA_J^b \calA_K^c \rp .
\end{align}
From Eqs.~\eqref{trans_del} and \eqref{trans_Ai}, the color magnetic field in terms of the conformal variables are expressed as
\begin{align}
	B_I^a = \epsilon_{IJK} \lp \tilde{\del}_J A_K^a  -\half \epsilon^{abc} A_J^b A_K^c \rp
	\label{mag} .
\end{align}
%
%
%}
We assume that the gauge field itself is also homogeneous and boost-invariant.
%\sout{The simplest}\comm{
One possible
%} 
realization of the gauge configuration
% \sout{to fulfill the above conditions and to exactly satisfy Gauss's law}
is given by
\begin{align}
	A^a_i &= \tilA(\theta) \lp \delta^{a1}\delta_{iy} + \delta^{a2}\delta_{ix} \rp,
	\label{bgconf} \\
	A^a_\eta &= 0\ ,
\end{align}
which fulfills the above requirements and exactly satisfies Gauss's law.
Taking the configuration Eq.~\eqref{bgconf}, 
only the second term in Eq.~\eqref{mag} remains. 
This is the generalization of the nonexpanding case~\cite{Berges12a,Tsutsui:2014rqa}.
We also mention that there exists not only color magnetic fields
but also transversely polarized color electric fields.

The EOM of the background field is now given by
\begin{align}
	\deltheta^2 \tilA + \frac{1}{4\theta^2}\tilA + \tilA^3 = 0. 
	\label{eombg}
\end{align}
The second term comes from the quadratic term in the action
which makes it difficult to get exact solutions analytically.
Nonetheless, the qualitative behavior is well understood 
in some limiting cases as we shall show.

In the earlier time $\theta \ll 1/\tilA$,
the self interaction term $\tilA^3$ can be neglected and the EOM is reduced to a linear equation.
Suppose that 
%\sout{at $\theta = \theta_0$,taking an}
the initial condition is given as 
$\tilA=\sqrt{B_0}$ and $\deltheta\tilA=0$
%\sout{for instance}
at $\theta = \theta_0$,
then the solution is given by
\begin{align}
	\tilA(\theta) \simeq \sqrt{\frac{B_0 \theta}{\theta_0}} \lp 1 - \half \log\frac{\theta}{\theta_0} \rp .
	\label{early_bg}
\end{align}

The nonlinear term dominates over the linear term in Eq.~\eqref{eombg}
after a time,
and the EOM becomes nonlinear but solvable at the later times $\theta \gg 1/\tilA$ where the linear term is negligible;
The background field should behave as
\begin{align}
	\tilA(\theta) 
	&\simeq 
%	\sqrt{B_\text{eff}}
	B_\text{eff}^{1/2}
%	\cn\lp\sqrt{B_\text{eff}}\theta+\Delta);1/\sqrt{2}\rp ,
	\cn\lp\theta{B_\text{eff}^{1/2}}+\Delta;1/\sqrt{2}\rp ,
	\label{bg_asy}
\end{align}
where $\cn(\theta;k)$ is the Jacobi elliptic function of modulus $k$,
which is a periodic function in time whose period is $T = 4K(k)$.
$K(k)$ is the complete elliptic integral of the first kind.
Specifically, $T \simeq 7.42$ for $k = 1/\sqrt{2}$.
This is nothing but the homogeneous background field discussed in \cite{Berges12a,Tsutsui:2014rqa}
in the nonexpanding geometry with a {\em phase} shift $\Delta$ quantifying the earlier time effects.

In our setup, ${B_\text{eff}^{1/2}}$ is the only dimensionful scale
and its strength depends on the initial condition.
It is convenient to introduce dimensionless quantities 
so as to normalize the final amplitude of the background gauge field in accordance with Eq.~\eqref{bg_asy};
that is, the final amplitude of $\tilA/\rtB$ is normalized to 
unity and the combination $\theta B_\text{eff}^{1/2}$ is the dimensionless time variable.

\begin{figure}[bth]
\PSfig{8.0cm}{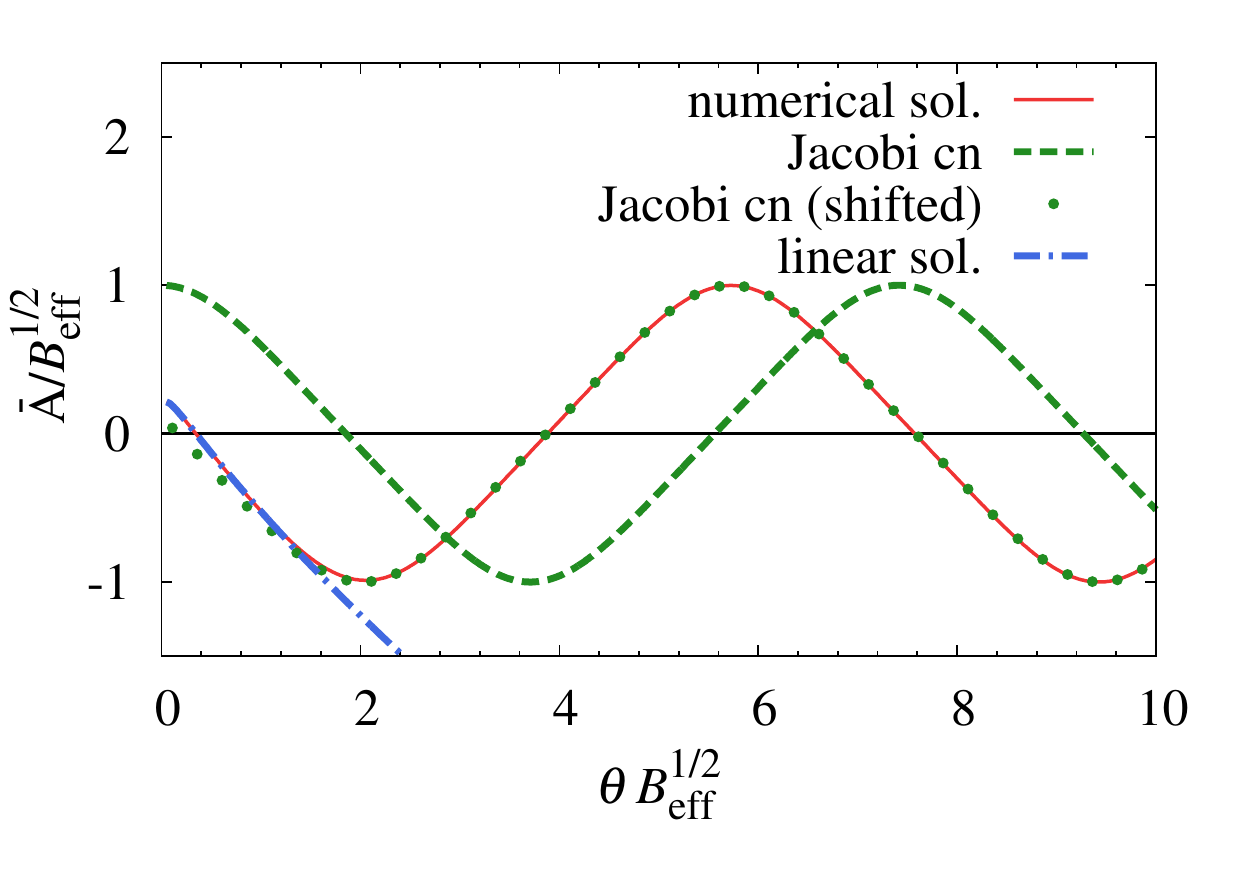}
	\caption{
Background field $\tilA$ in an expanding geometry. 
We show the numerical solution of the EOM Eq.~\eqref{eombg} rescaled by the asymptotic amplitude (red solid line), the linear solution Eq.~\eqref{early_bg} (blue dash-dotted line), the shifted Jacobi elliptic function Eq.~\protect\eqref{bg_asy} (green dotted line) and the Jacobi elliptic function (green dashed line).
The temporal variable $\theta$ and the amplitude $\tilA$ are normalized by the asymptotic amplitude $\rtB$.
	}
	\label{Fig:bg}
\end{figure}

Figure~\ref{Fig:bg} shows the numerical solution of the EOM Eq.~\eqref{eombg} rescaled by the final amplitude $B_\text{eff}^{1/2}$.
We also show the solution of the linear equation Eq.~\eqref{early_bg}, the Jacobi elliptic function, and the shifted Jacobi elliptic function Eq.~\eqref{bg_asy}.
At the initial time,
the initial gauge field and its derivative are chosen to be 
$\theta_0=0.01$, $\tilde{A}(\theta_0)=1$ and $d\tilde{A}/d\theta(\theta_0)=0$,
respectively.
The final amplitude and the time shift are found to be $B_\text{eff}^{1/2}\simeq4.68$ and $\Delta\simeq1.69$.
Thus the numerical result in Fig.~\ref{Fig:bg} starts from $\theta_0 B_\text{eff}^{1/2} \simeq 0.01\times 4.68$.

The linear solution Eq.~\eqref{early_bg} 
gets to fail to reproduce the numerical solution 
around $\theta B_\text{eff}^{1/2}\sim1$ as expected.
Let us compare this time with the transition time from the linear to the nonlinear regime $\theta_\text{tr}$ obtained from the balance condition $1/\theta_\text{tr} \sim \tilA(\theta_\text{tr})$.
Ignoring the logarithmic correction,
we get 
$\theta_\text{tr} \sim (\theta_0/B_0)^{1/3}$ from Eq.~\eqref{early_bg}.
By assuming $\theta_\text{tr}B_\text{eff}^{1/2}\sim1$,
we find $B_\text{eff}\sim(B_0/\theta_0)^{1/3}=(1/0.01)^{1/3}\simeq4.64$
for the present initial condition,
which deviates from the numerical results only by around 1\%.
%
%
%\begin{align}
%	\tilA(\theta) 
%	&\simeq 
%	\lp\frac{B_0}{\theta_0}\rp^{1/3} \cn\lbb\lp\frac{B_0}{\theta_0}\rp^{1/3}\theta;1/\sqrt{2}\rbb \\
%	&\equiv 
%	\sqrt{B_\text{eff}} \cn\lp\sqrt{B_\text{eff}}\theta;1/\sqrt{2}\rp .
%	\label{bg_asyA}
%\end{align}
%The final amplitude $\sqrt{B_\text{eff}}$ diverges if the starting time is chosen to be $\theta_0 \rightarrow 0$.

%%

Before finishing this subsection,
we comment on the physical time scale.
From Eqs.~\eqref{trans_Ai}, \eqref{confornal_time}, 
	\eqref{early_bg}, and~\eqref{bg_asy},
the solutions at the earlier and later times
in the original $\tau$-$\eta$ coordinate are given by
\begin{align}
%	&\tilde{\calA}(\tau) 
	%\equiv 
	&\calA^1_y(\tau)
	= \calA^2_x(\tau)
	= \tau^{-1/3} \tilde{A}(\theta)
%	\lp\frac{B_0}{\theta_0\tau}\rp^{1/3} 
%	\cn\lbb \frac{3}{2}\lp\frac{B_0\tau^2}{\theta_0}\rp^{1/3} + \lp B_0 \theta_0^2 ;1/\sqrt{2}\rp \rbb 
\nonumber\\
	&\simeq
\begin{cases}
	{\displaystyle
	\sqrt{\frac32} B_\mathrm{eff}^{3/4} 
	\left[1-\frac13\log\left(\frac{\tau}{\tau_0}\right)\right]}
%	&\mkern-55mu(\theta\rtB\ll 1)\ ,\\
	&(\theta\rtB\ll 1)\ ,\\
	\\
	{\displaystyle
	\frac{\rtB}{\tau^{1/3}} \cn\lp 
	\frac32\tau^{2/3}\rtB+\Delta ;\frac{1}{\sqrt{2}}\rp }
% \\
%	&\mkern-55mu(\theta\rtB\gg 1)\ .
	&(\theta\rtB\gg 1)\ ,
\end{cases}
	\label{bg_asy_taueta}
%	&\equiv
%	\calB_\text{eff}^{1/3}\tau^{-1/3} \cn\lp 3\calB_\text{eff}^{1/3}\tau^{2/3}/2 ;1/\sqrt{2}\rp .
\end{align}
respectively.
Equation~\eqref{bg_asy_taueta} tells us the strength of the background gauge field.
If the initial strength at $\tau_0=(2\theta_0/3)^{3/2}$
is given by the saturation scale $Q_\text{s}$,
we get the relation between the physical and conformal scales,
\begin{align}
	Q_\mathrm{s}=&\sqrt{\frac32}\,B_\mathrm{eff}^{3/4}\ ,\quad
	Q_\mathrm{s}\tau=\frac23\, \left(\theta \rtB\right)^{3/2}\ .
\label{Eq:timescale}
\end{align}
%\comm{
%	the proper time $\tau$ is expressed in terms of the dimensionless conformal time $\theta B_\text{eff}^{1/2}$ by $\tau = (2\theta B_\text{eff}^{1/2}/3)^{3/2} \times (2\theta_0 B_\text{eff}^{1/2}/3)^{-1/2} \times Q_\text{s}$.
%	Here, we use the relation $\tau_0^{2/3}=2\theta_0/3$.
%}
%\comm{
%	Specifically, in our calculation,  
%	$\tau \simeq (2\theta B_\text{eff}^{1/2}/3)^{3/2} \times 5.66 Q_\text{s}$
%	by putting the initial time $\theta_0 B_\text{eff}^{1/2} \simeq 0.01\times 4.68$.
%}
For instance, we can evaluate the proper time as
$Q_\mathrm{s}\tau = 0.67, 3.46$ and $21.1$
for
$\theta\rtB=1, 3$ and $10$, respectively.

%%%%%%%%%%%%%%%%%%%%%%%%%%%%%%%%%%%%%%%%%%%%%%%%%%%%%%%%%%%%%%
\subsection{Equation of motion of fluctuations}\label{subsec:fluc}
We write down the linearized EOM of fluctuations on top of the background gauge field Eq.~\eqref{bgconf}.
This is easily done by shifting $A^a_I \rightarrow A^a_I + a^a_I$ in Eqs.~\eqref{eom_i} and \eqref{eom_eta}, and keep terms of the order $\mathcal{O}(a^{1})$.
The resultant EOM of fluctuations $a^a_I$ is given by
\begin{align}
	&\deltheta^2 a^a_i + \frac{1}{9\tiltheta^2}a^a_i 
	+ \lbb \Omega^2(\tilA) \rbb^{ab}_{ij} a^b_j 
	+ \lbb \Omega^2(\tilA) \rbb^{ab}_{i\eta} a^b_\eta  = 0 \ ,
	\label{linearEOMt}
\\
	&\frac{1}{\tiltheta^3} \calL^2_\eta a^a_\eta + \frac{7}{9\tiltheta^5}a^a_\eta 
	+ \lbb \Omega^2(\tilA) \rbb^{ab}_{\eta j} a^b_j 
	+ \lbb \Omega^2(\tilA) \rbb^{ab}_{\eta\eta} a^b_\eta  = 0
	\label{linearEOM}
\end{align}
where $\calL^2_\eta$ and $\Omega^2_I$ are defined by 
\newlength{\len}
\settowidth{\len}{=\,\,\,}
%\the\len
\begin{align}
	\calL^2_\eta =& \frac{d^2}{d\theta^2} - \frac{2}{\tiltheta}\frac{d}{d\theta},
\end{align}
and
\begin{align}
	%%
	%%
%	M^2_I &= \frac{1}{9\tiltheta^2}\delta_{Ii} + \frac{7}{9\tiltheta^5} \delta_{I\eta}, \\
%	\calM^2_i &= M^2_i = \frac{1}{9\tiltheta^2}\ ,\quad
%	\calM^2_\eta = \tiltheta^{-3}M^2_\eta = \frac{7}{9\tiltheta^5}\ , \\
	%%
	%%
%	\lbb \Omega^2 \rbb^{ab}_{IJ} 
%	&=
%	\delta_{Ii}\delta_{Jj} \lbb 
%	-\lp \tilde{D}_k\tilde{D}_k + \frac{1}{\tiltheta^3} \tilde{D}_\eta \tilde{D}_\eta \rp\delta_{ij}
%	+ \tilde{D}_i\tilde{D}_j + 2i\tilde{F}_{ij} \rbb \notag \\
%	%
%	&\hspace{\len}
%	+\delta_{Ii}\delta_{J\eta}
%	\frac{1}{\tiltheta^3} \lp \tilde{D}_i\tilde{D}_\eta + 2i\tilde{F}_{i\eta} \rp \notag \\
%	%
%	&\hspace{\len}
%	+\delta_{I\eta}\delta_{Jj}
%	\frac{1}{\tiltheta^3} \lp \tilde{D}_\eta \tilde{D}_j + 2i\tilde{F}_{\eta j} \rp \notag \\
%	%
%	&\hspace{\len}
%	-\delta_{I\eta}\delta_{J\eta}
%	\frac{1}{\tiltheta^3} \tilde{D}_k \tilde{D}_k.
%\\
	%% ij
	\lbb \Omega^2 \rbb^{ab}_{ij} 
	=&
	-\lp \tilde{D}_k(\tilA)\tilde{D}_k(\tilA) + \frac{1}{\tiltheta^3} \tilde{D}_\eta(\tilA) \tilde{D}_\eta(\tilA) \rp^{ab}\delta_{ij} \nonumber\\
	&+ \left(\tilde{D}_i(\tilA)\tilde{D}_j(\tilA) + 2i\tilde{F}_{ij}(\tilA)
	\right)^{ab} \ ,  \\
	\lbb \Omega^2 \rbb^{ab}_{i\eta} 
	=&\frac{1}{\tiltheta^3} \lp \tilde{D}_i(\tilA)\tilde{D}_\eta(\tilA) + 2i\tilde{F}_{i\eta}(\tilA) \rp^{ab} \ ,\\
	\lbb \Omega^2 \rbb^{ab}_{\eta{j}} 
	=&\frac{1}{\tiltheta^3} \lp \tilde{D}_\eta(\tilA) \tilde{D}_j(\tilA) + 2i\tilde{F}_{\eta j}(\tilA) \rp^{ab} \ ,\\
	\lbb \Omega^2 \rbb^{ab}_{\eta{j}} 
	=&-\frac{1}{\tiltheta^3} \lp\tilde{D}_k(\tilA) \tilde{D}_k(\tilA)
	\rp^{ab} \ ,
\end{align}
respectively.
The color indices $ab$ appear in the covariant derivative $\tilde{D}$ and 
the field tensor $\tilde{F}$.
$\Omega^2$ depends on conformal time not only explicitly but also implicitly through the background gauge field $\tilA(\theta)$.
Here and in the later discussions, we adopt the unit $\rtB=1$.
Namely, all variables are normalized by $\rtB$ such as $\theta\rtB$ and $a_I^a/\rtB$.

Even though we consider the color magnetic background,
both the transverse momenta $p_i$ and the longitudinal momentum $p_\eta$ are well defined because the background gauge field does not depend on spatial coordinates.
Therefore, it is useful to introduce the Fourier representation of the EOM, Eqs.~\eqref{linearEOMt} and \eqref{linearEOM} through
\begin{align}
	a^a_I(\theta,x,y,\eta) = \int \frac{d^3p}{(2\pi)^3} a^a_I(\theta,p_x,p_y,p_\eta) e^{i(p_x x + p_y y + p_\eta \eta)} .
\end{align}
It should be noted that we can set $p_y = 0$ without loss of generality 
from rotational symmetry in the transverse plane.
The matrix $\Omega^2$ at $p_y=0$ becomes block-diagonal as $\Omega^2 = \diag\lp \Omega_4^2,\, \Omega_5^2  \rp$, as found in the nonexpanding case~\cite{Tsutsui:2014rqa}.

In the later discussion, we consider finite $p_\eta$ modes
$(p_\eta\neq0, p_T=0)$ and finite $p_T$ modes $(p_\eta=0, p_T\neq0)$.
From now on, $p_T$ denotes a transverse momentum.
The EOM of these modes at later times have the form of
\begin{align}
	\frac{d^2a^a_I}{d\theta^2}  + k^2_\text{eff}(\theta) a^a_I 
	+ \lambda\, \cn^2(\theta+\Delta) a^a_I = \text{inhomogeneous terms} , \label{typical_form}
\end{align}	
and $d^2/d\theta^2$ is replaced with $\calL^2_\eta$ for the longitudinal component $a^a_\eta$. 
Here, we defined the effective momenta $k^2_\text{eff}(\theta)$ for convenience.
The effective momenta for the transverse and longitudinal components of the finite $p_\eta$ (finite $p_T$) modes,
$k^2_\text{eff}=k^2_{\eta T}$ and $k^2_{\eta\eta}$ 
($k^2_{TT}$ and $k^2_{T\eta}$),
are defined by Eqs.~\eqref{keT}-\eqref{kTe} below.
The explicit forms of the EOMs are summarized in Appendix A.

For finite $p_\eta$ modes ($p_T=0$), 
we find the effective momenta of the transverse and longitudinal components of fluctuations $a^a_i$ and $a^a_\eta$ given by
\begin{align}
	k^2_{\eta T}(\theta) &\equiv M^2_T(\theta) + \frac{p^2_\eta}{\tiltheta^2} = \frac{1+9p^2_\eta}{4\theta^2}, \label{keT} \\
	k^2_{\eta \eta}(\theta) &\equiv M^2_\eta(\theta) = \frac{7}{4\theta^2},
	\label{kee}
\end{align}
where $M^2_T$ and $M^2_\eta$ are effective masses defined by
\begin{align}
	M^2_T(\theta) = \frac{1}{9\tiltheta^2} = \frac{1}{4\theta^2} ,\quad
	M^2_\eta(\theta) = \frac{7}{9\tiltheta^2} = \frac{7}{4\theta^2}.
	\label{effmass}
\end{align}
Note that $k^2_{\eta T}$ and $k^2_{\eta\eta}$ decrease in time monotonically.

In the same manner, 
we define the effective momenta for finite $p_T$ modes ($p_\eta=0$) by 
\begin{align}
	k^2_{TT}(\theta) &\equiv M^2_T(\theta) + \tiltheta p_T^2 = \frac{1}{4\theta^2} + \frac{2}{3}\theta p_T^2, \label{kTT}\\
	k^2_{T\eta}(\theta) &\equiv M^2_\eta(\theta) + \tiltheta p_T^2 = \frac{7}{4\theta^2} + \frac{2}{3}\theta p_T^2. \label{kTe}
\end{align}
In contrast to the finite $p_\eta$ modes Eqs.~\eqref{keT} and~\eqref{kee}, $k^2_{TT}$ and $k^2_{T\eta}$ increase at later times.

Before we close this section,
we comment on the similarities and differences in EOMs in the expanding and nonexpanding geometries.
The forms of EOMs expressed by using the conformal variables Eqs.~\eqref{linearEOMt} and~\eqref{linearEOM} are similar to those in a nonexpanding geometry.
For example,
the background field at later times is described by the elliptic function, 
then the EOMs in Eq.~\eqref{typical_form} is similar to Lam\'{e}'s equation, 
which appear in the nonexpanding geometries and discussed in the next section.
In addition,
$\Omega^2$ is block-diagonalized in the same way as in the nonexpanding case~\cite{Tsutsui:2014rqa}.
On the other hand,
EOMs in the expanding geometry is different from those in the nonexpanding geometry in the appearance of the time-dependence of the effective transverse and longitudinal momenta.
Thus, the original problem defined in the expanding geometry is mapped into the nonexpanding problem where the momenta of the fluctuations depend on time.

%%%%%%%%%%%%
%%%%%%%%%%%%%%%%%%%%%%%%%%%%%%%%%%%%%%%%%%%%%%%%%%%%%%%%%%%%%%
%\section{Overview of Floquet theory}\label{sec:Floquet}
\section{Floquet theory}\label{sec:Floquet}
In general, a temporally periodic-driven system can show instability due to the resonance between an external force and eigenmodes of the system.
The resultant instability is called parametric resonance or parametric instability.
It is well known that the Floquet theory is best suited to analyze instabilities of temporally periodic-driven systems.
Floquet theory provides the criteria whether the solution is exponentially, polynomially divergent or bounded.
In the expanding problem,
Floquet theory is not directly applied but 
tells us how the fluctuations behave asymptotically. 
An account of the Floquet theory is given in our previous paper~\cite{Tsutsui:2014rqa}.
In this section,
we recapitulate them to be self-contained in a brief way.
More complete discussions and applications, see also Appendix~\ref{App:Floquet}.

Let us consider Lam\'{e}'s equation
\begin{align}
	\ddot{f} + \lbb p^2 + \lambda \cn^2(t;k) \rbb f = 0 .
	\label{pLame}
\end{align}
Here, $p$ and $\lambda$ are control parameters of this equation.
It should be noted that, at later times, EOMs in the expanding geometry, Eq.~\eqref{typical_form}, becomes the Lam\'{e}'s equation if we ignore the time-dependence of the effective momenta and the inhomogeneous terms.
 In the context of the CYM theory, the cases with
$p\ll1$ and $\lambda = \pm1, 2, 3$ are important.
Let us define a fundamental matrix by
$\Phi(t)=((f_1, \dot{f_1})^t, (f_2, \dot{f_2})^t)$,
where $\{f_i\}_{(i=1,2)}$ are independent and complete solutions of the Lam\'{e}'s equation. 
If $\Phi(t)$ is a fundamental matrix, 
or in this specific case, a Wronskian matrix, 
$\Phi(t+T)$ is also a fundamental matrix due to the periodicity of the elliptic function,
where $T$ is the period of the elliptic function.

The criterion of the existence of unstable solutions is expressed in terms of the monodromy matrix
$M$ defined by $\Phi(t+T) = \Phi(t) M$ 
which is regular and time-independent.
By construction, 
the fundamental matrix is also regular and we get
\begin{align}
	M = \Phi(0)^{-1} \Phi(T) .
\end{align}	
Because the Wronskian $\det \Phi$ is constant in time,
we also find $\det M=\mu_1\mu_2=1$.
The eigenvalues of $M$ are called characteristic multipliers and we denote them as $\mu_1$ and $\mu_2$.
These multipliers are the solutions of the characteristic equation: $\mu^2 - (\tr M)\mu + 1 =0$.

\begin{figure}[bth]
	\PSfig{8.0cm}{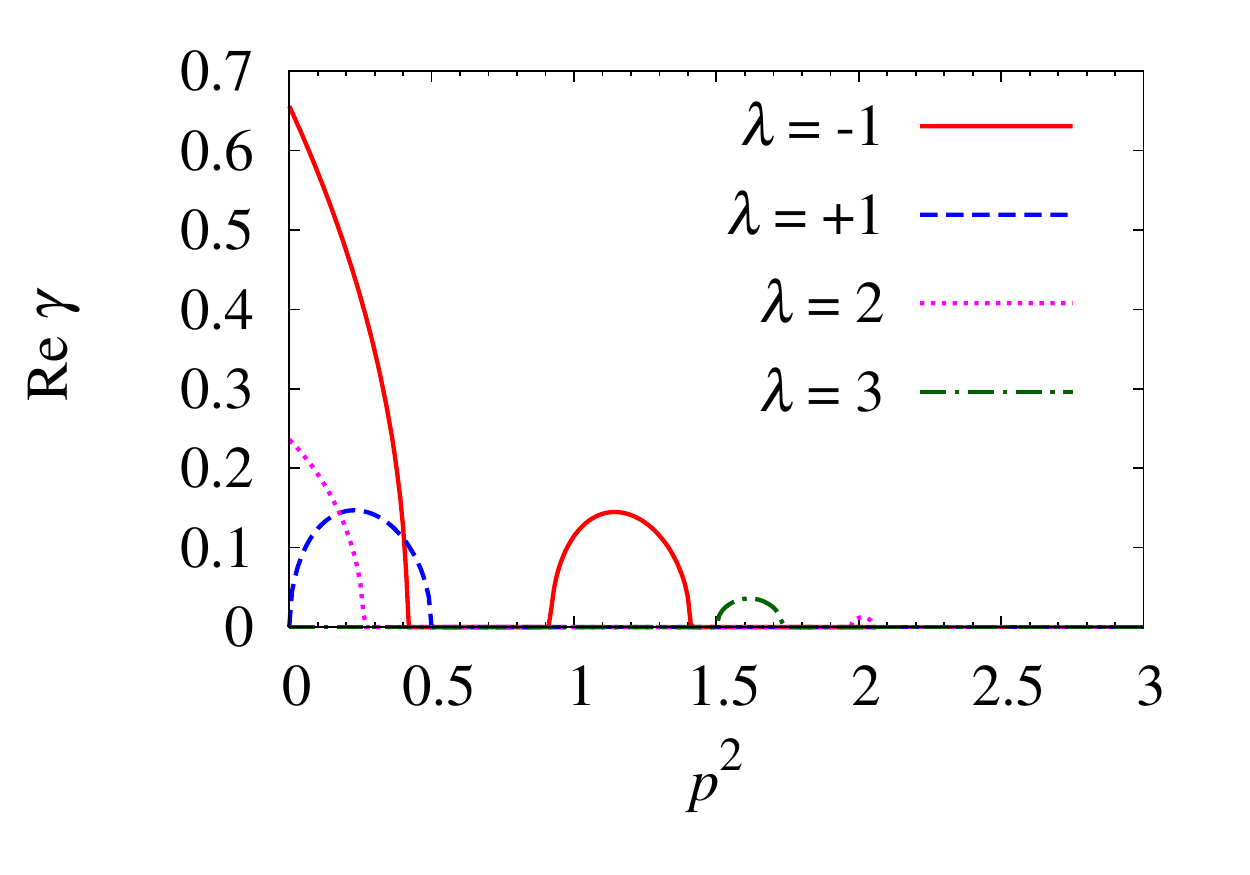}
	\caption{
		Growth rates in Lam\'e's equation. 
		We show the real part of the growth rates for 
		$\lambda =-1$ (red solid line), 
		$\lambda =1$ (blue dashed line), 
		$\lambda =2$ (magenta dotted line) and
		$\lambda =3$ (green dot-dashed line). 
	}
	\label{Fig:lames}
\end{figure}

One can easily show that the solution of the Lam\'e's equation must have the following form:
\begin{align}
	\Phi(t) = F(t)\exp\lbb (\log M)\frac{t}{T}\rbb ,
	\label{Floquet thm}
\end{align}
where $F(t)$ is a $T$-periodic matrix whose specific form is irrelevant to our discussion.
It is useful to define the characteristic exponent or growth rate by $\gamma = (\log\mu)/T$.
Thus, the eigenvalues of the monodromy matrix determine the stability of the solution.
The complete classification is listed below:
\begin{enumerate}
	\item If $|\tr M| > 2$, the solution is exponentially divergent. 
	\item If $|\tr M| = 2$, the solution is (anti)periodic or linearly divergent. 
	\item If $|\tr M| < 2$, the solution is bounded. 
\end{enumerate}
Figure~\ref{Fig:lames} shows that the real part of the growth rates of Lam\'{e}'s equation for $\lambda = \pm1,2$ and $3$.
The growth rate has a maximum at $p^2=0$ for $\lambda=-1$ and $2$, whose values are $0.66$ and $0.23$, respectively.
For $\lambda=1$ and $3$,
the exponential growth rates are equal to zero at $p=0$
and the solutions are bounded or diverge at most linearly~\cite{Tsutsui:2014rqa}.
%%
%

%%%%%%%%%%%%
%%%%%%%%%%%%%%%%%%%%%%%%%%%%%%%%%%%%%%%%%%%%%%%%%%%%%%%%%%%%%%
\section{Stability analysis}\label{sec:Stability}

In this section,
we perform a linear stability analysis for fluctuations 
by solving the EOMs, Eqs.~\eqref{linearEOMt} and \eqref{linearEOM},
around the background field~\eqref{bgconf}.
We first show the numerical results in Sec.~\ref{subsec:num},
and then in the following subsections,
we shall give a semi-analytical analysis in some limits, 
which helps to understand the numerical results.
We shall only show the results for finite $p_\eta$ modes ($p_\eta \neq 0$, $p_T = 0$) 
and finite $p_T$ modes ($p_\eta = 0$, $p_T \neq 0$),
which should be sufficient to see how instabilities emerge in an expanding system.
The details including all explicit expressions of the equations are presented in Appendix~\ref{App:detail}.

%(NOTE:the list of mass dimensions
%$[\theta]=-2/3$, $[p_\eta]=0$, $[p_T]=1$, $[\tilp_\eta]=-1/3$, $[\tilp_T]=2/3$, $[\tilA]=2/3$, $[\tiltheta^{-3}]=2$.
%$[\Omega^2_{TT}]=4/3$, $[\Omega^2_{\eta T}]=7/3$, $[\Omega^2_{\eta\eta}]=10/3$.
%)

%
%%%%%%%%%%%%%%%%%%%%%%%%%%%%%%%%%%%%%%%%%%%%%%%%%%%%%%%%%%%%%
\begin{figure*}[bth]
	\PSfig{5.0cm}{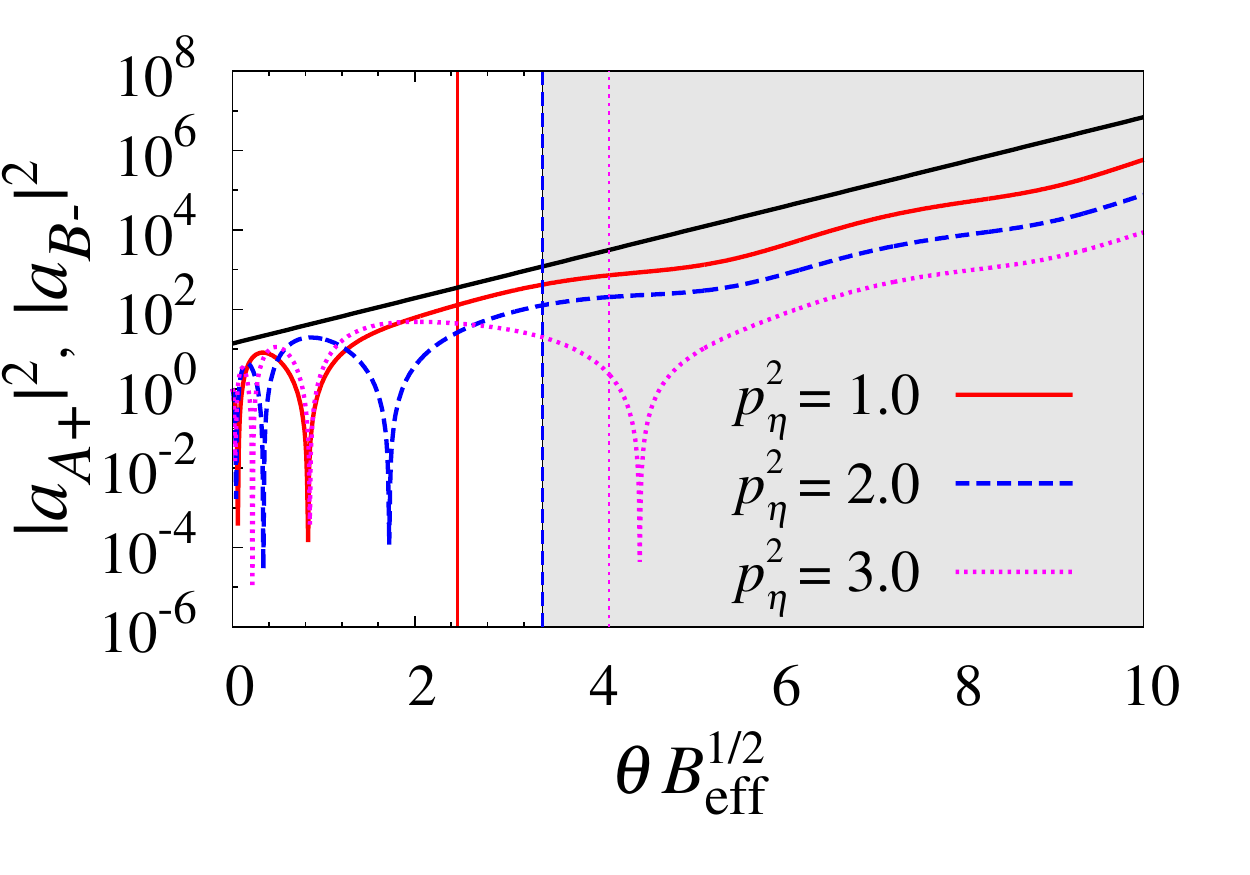}
	\PSfig{5.0cm}{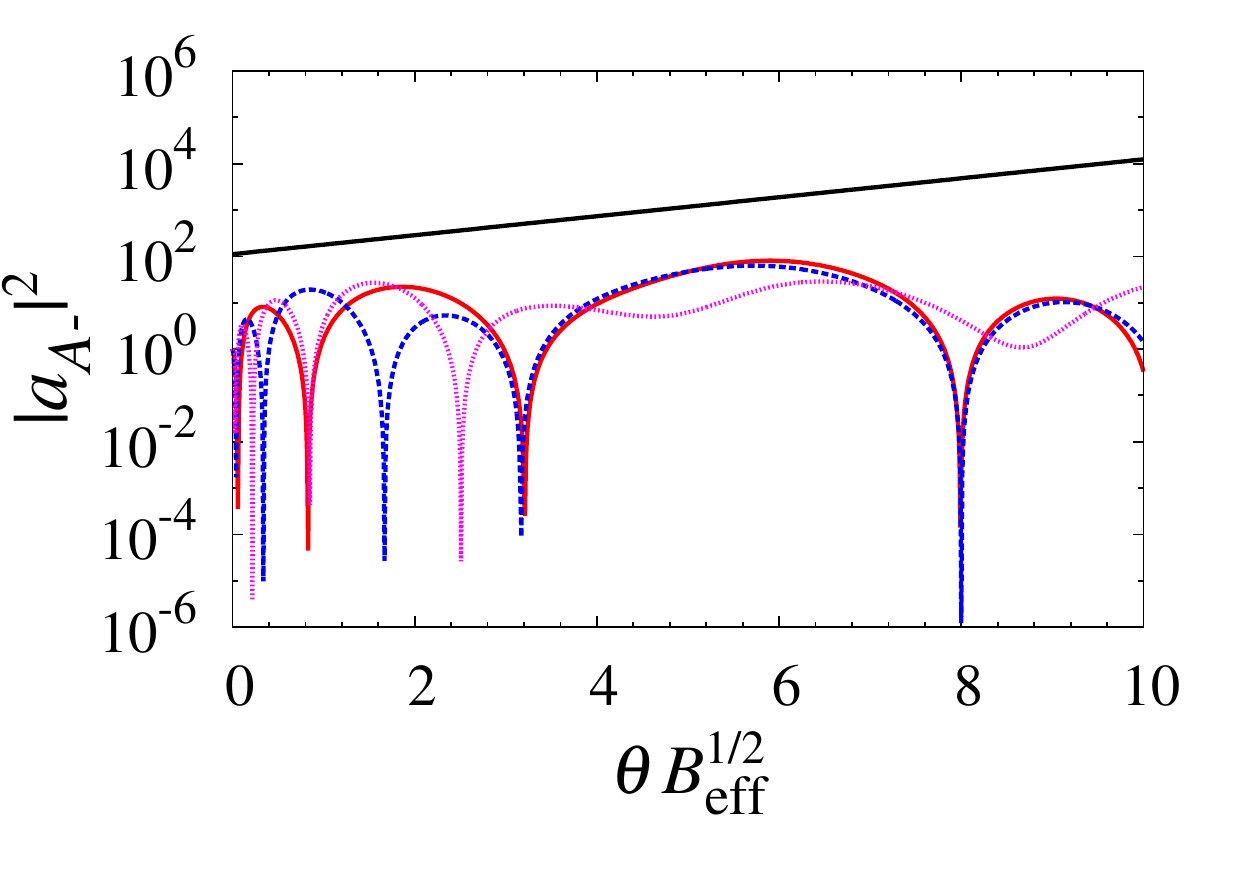}
	\PSfig{5.0cm}{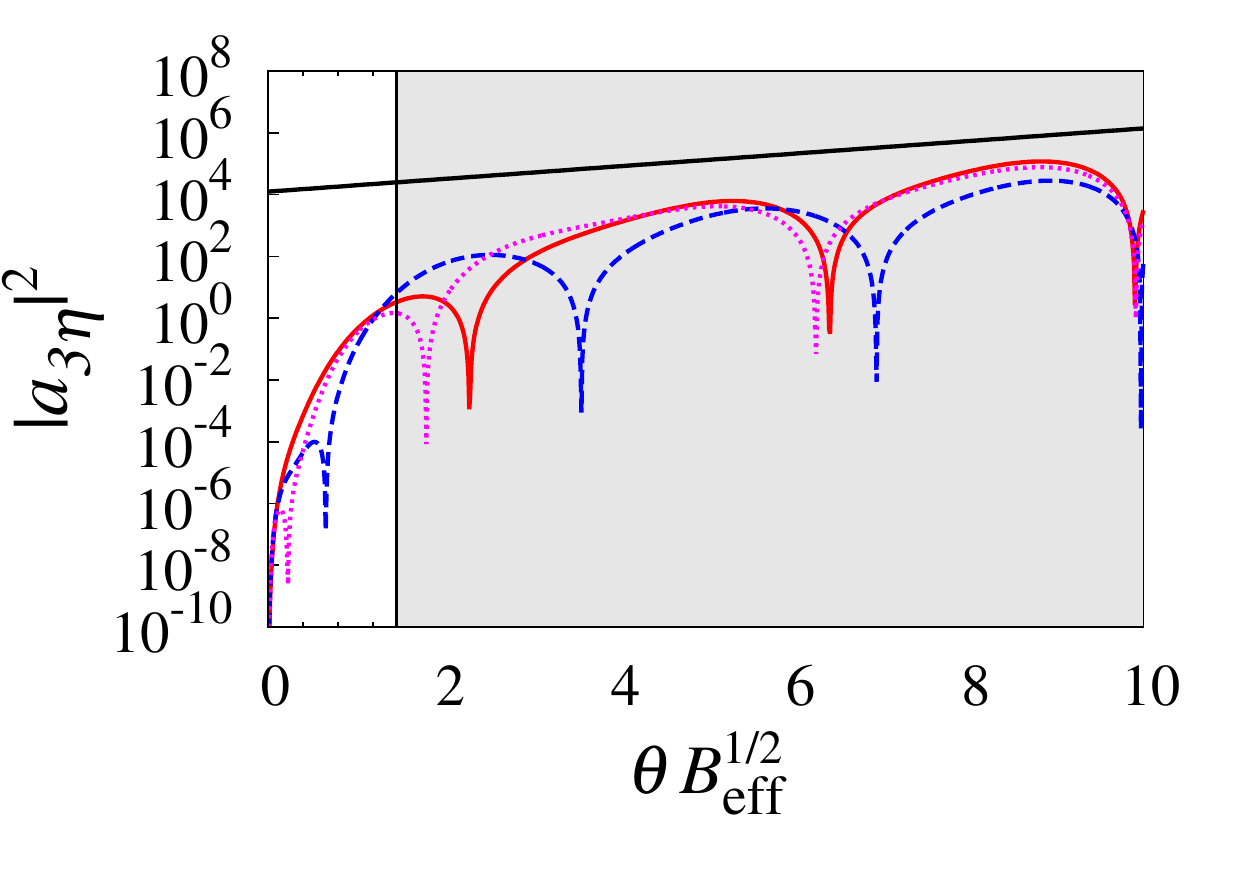}
	\PSfig{5.0cm}{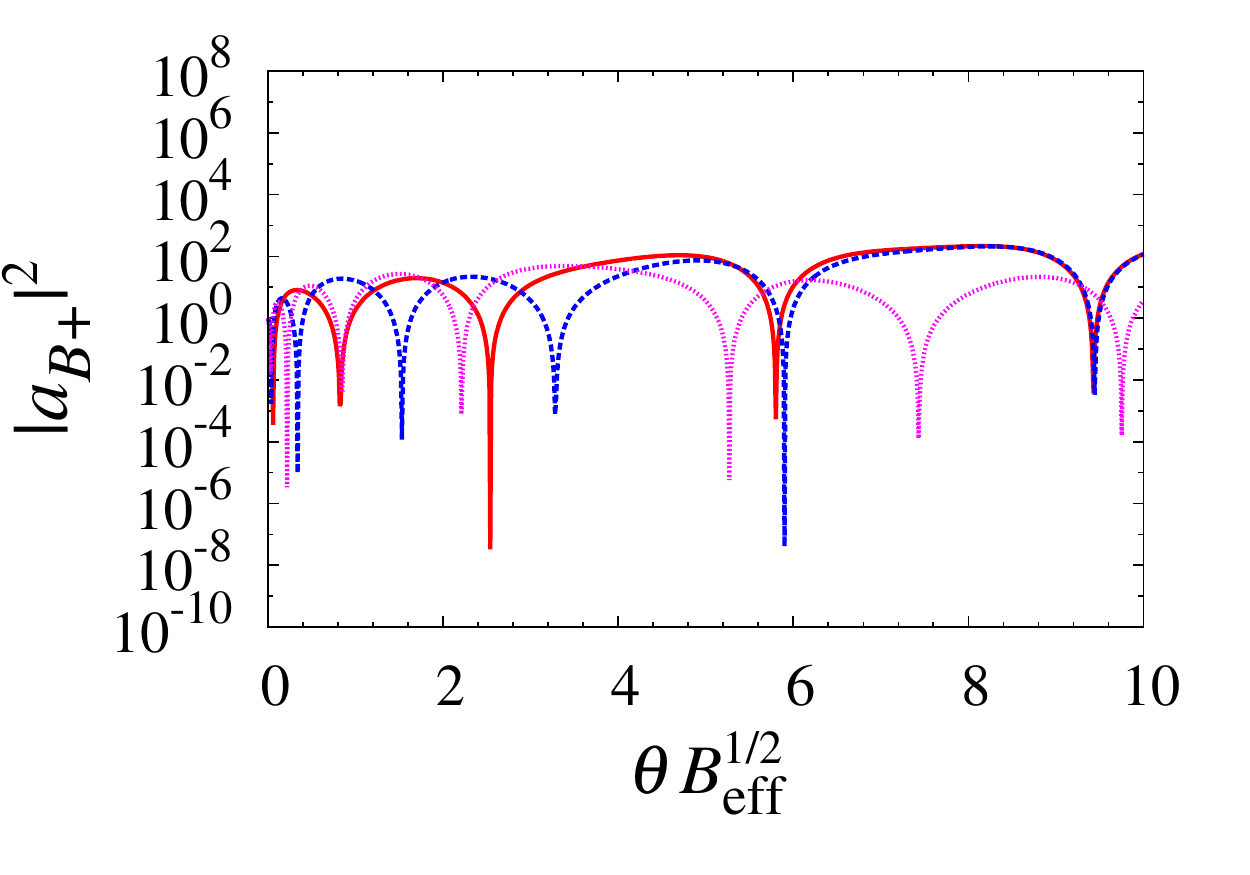}
	\PSfig{5.0cm}{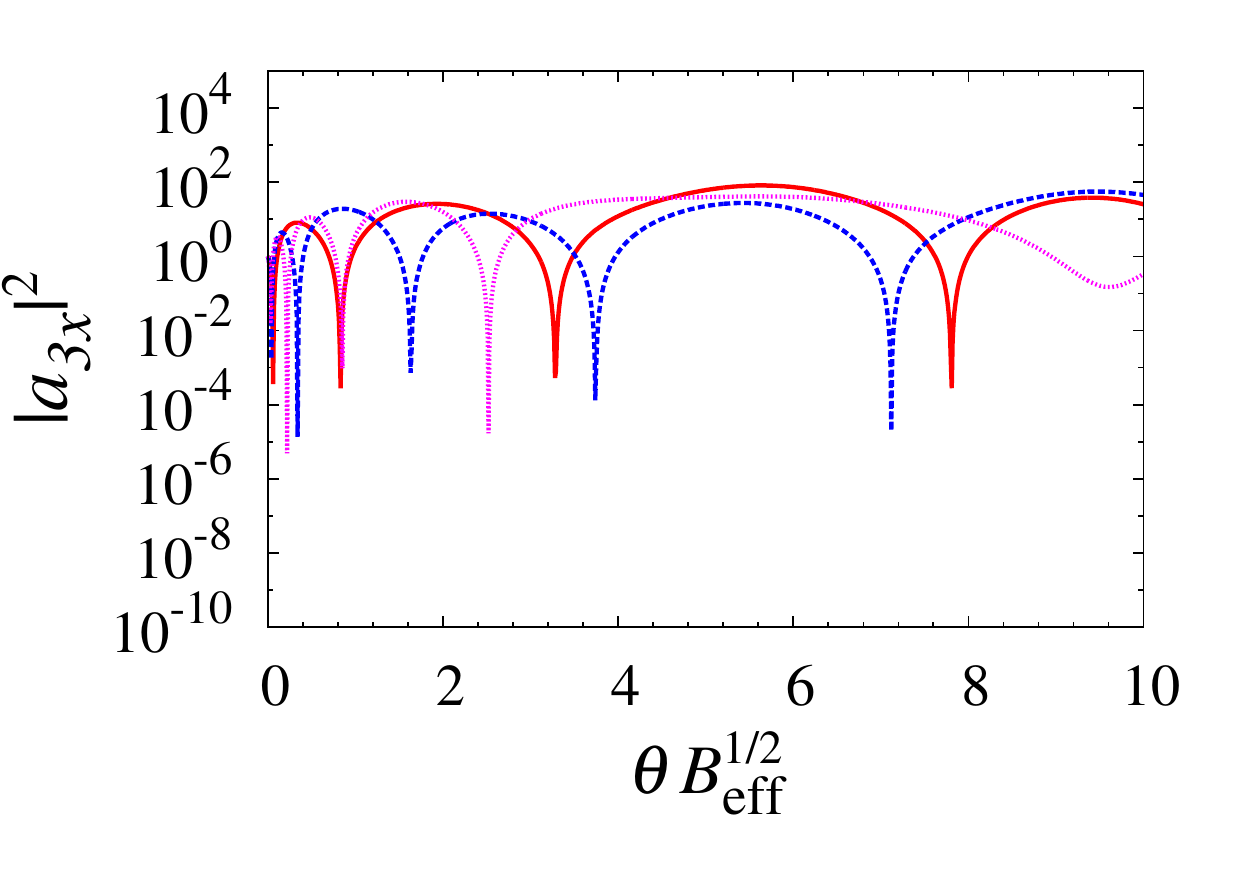}
	\PSfig{5.0cm}{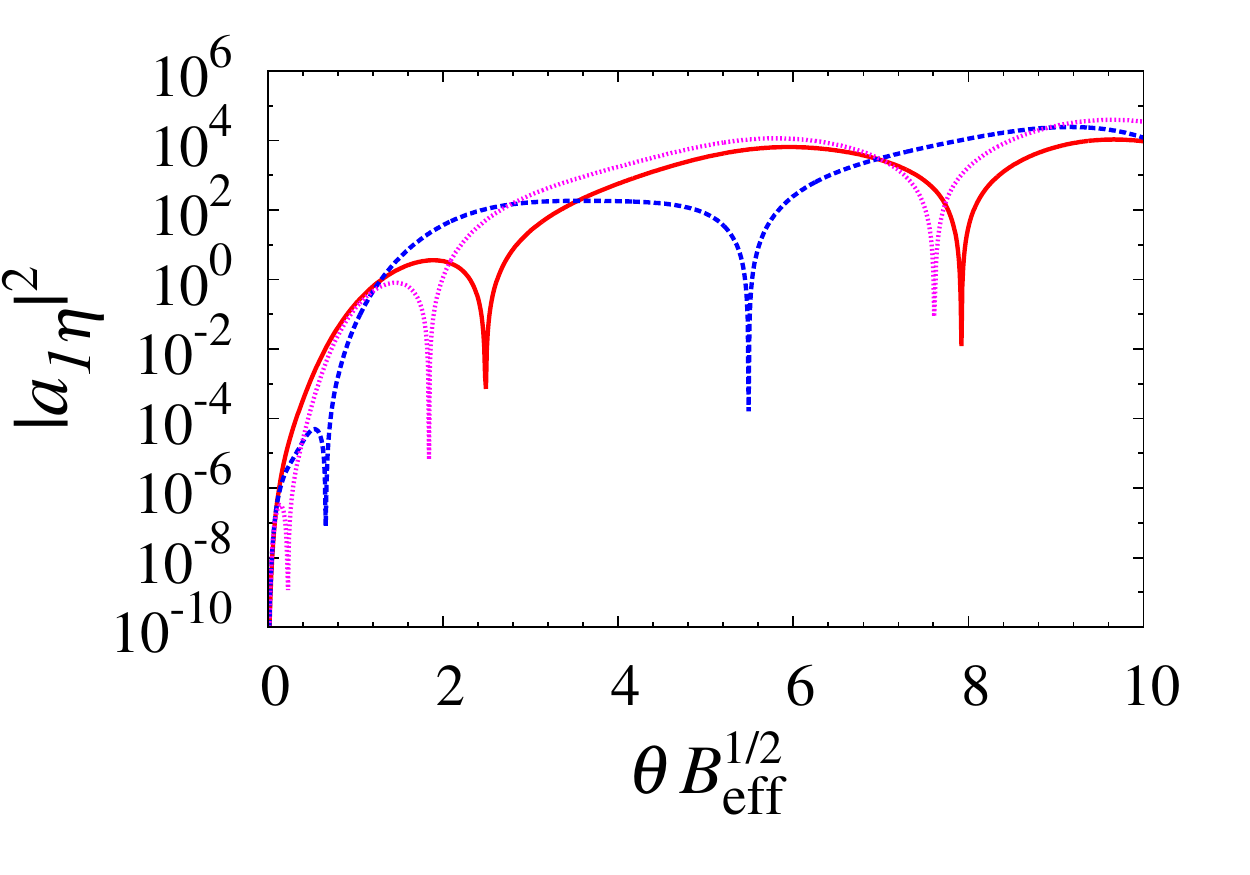}
	\caption{
	Squared fluctuation amplitudes in finite $p_\eta$ modes. 
	We show the time evolution of fluctuations in finite $p_\eta$ modes for $p_\eta^2 = 1.0, 2.0$ and $3.0$.
	Black solid lines in the upper panels show exponential functions $e^{2\gamma\theta}$
	with $\gamma=0.66$ ($a_{A+}$ and $a_{B-}$) 
	and $\gamma=0.23$ ($a_{A-}$ and $a^3_\eta$).
	The vertical lines show the transition times for each momentum mode.
	The transition times of $a_{A+}, a_{B-}$ are given by 2.46, 3.40 and 4.13 for each momentum mode.
	The transition time of $a^3_\eta$ is 1.97 and independent of $p_\eta$.
	Shaded areas show the time regime where exponential growth is expected for $p_\eta^2=2.0$.
	}
	\label{Fig:pL_trans}
\end{figure*}
\begin{figure*}[bth]
	\PSfig{5.0cm}{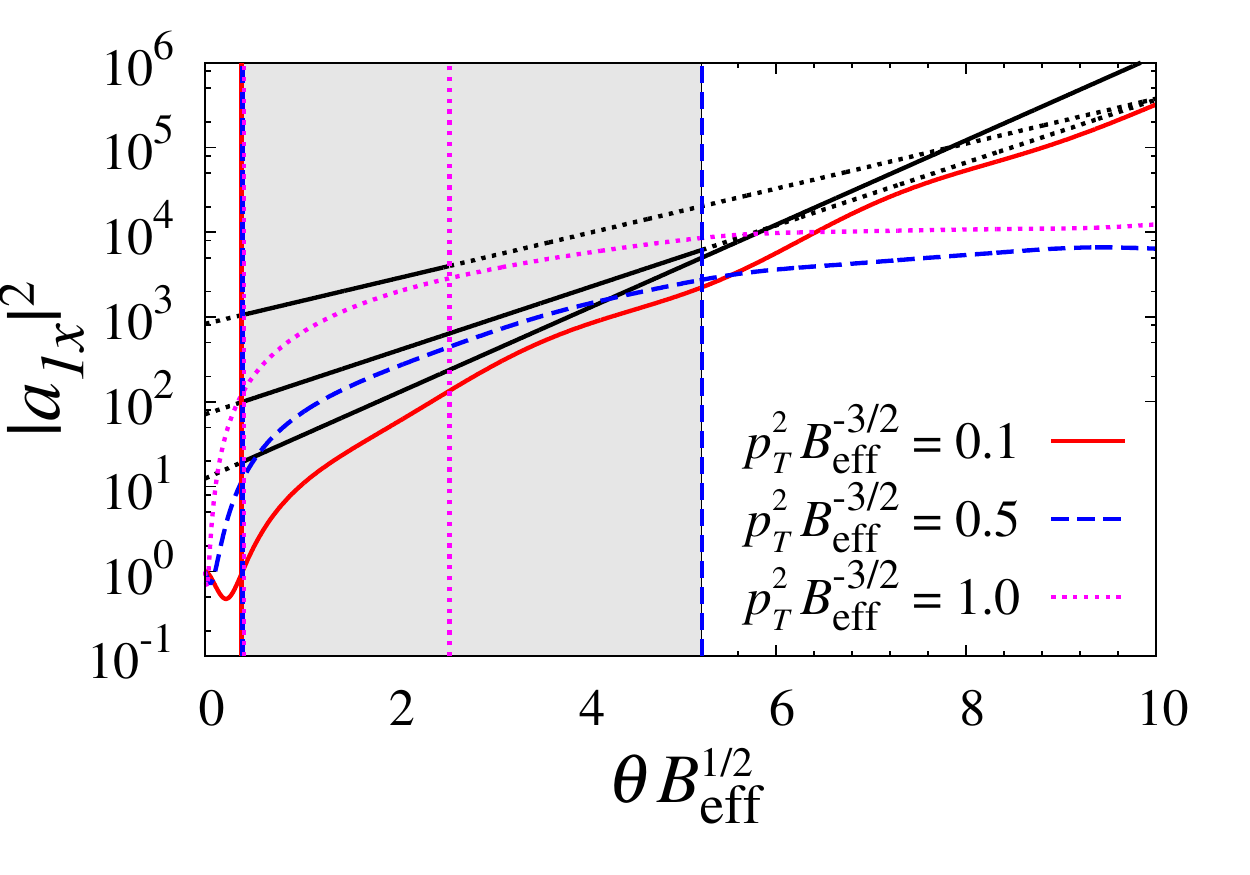}
	\PSfig{5.0cm}{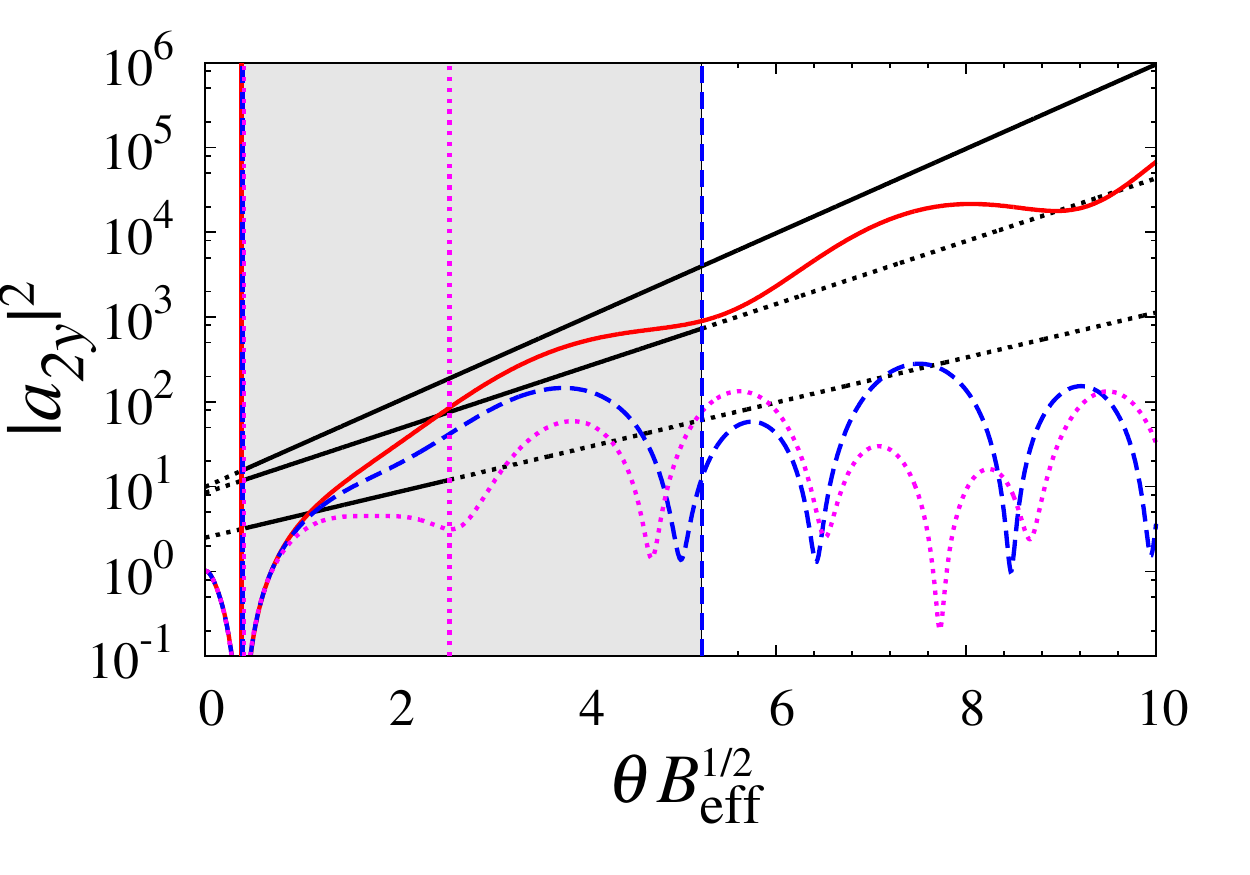}
	\PSfig{5.0cm}{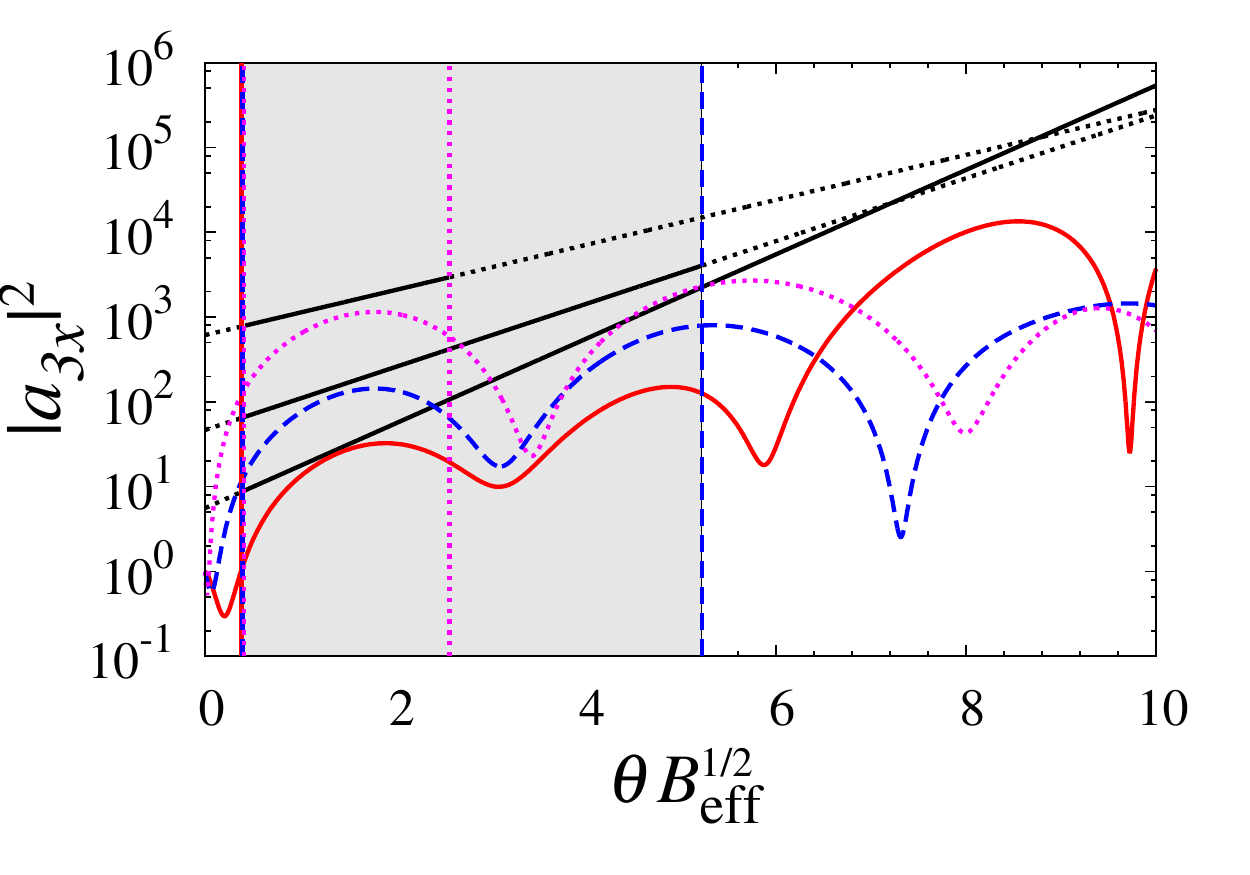}
	\PSfig{5.0cm}{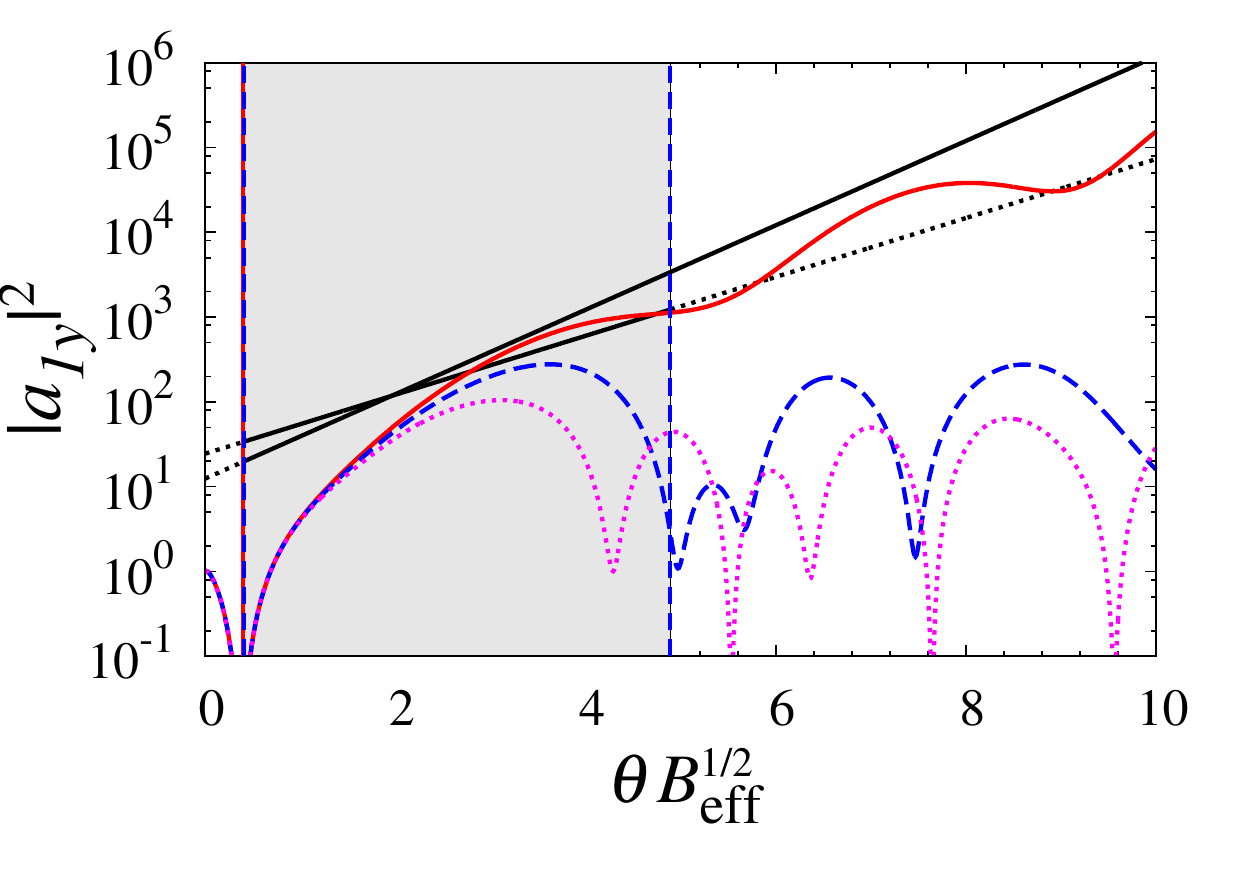}
	\PSfig{5.0cm}{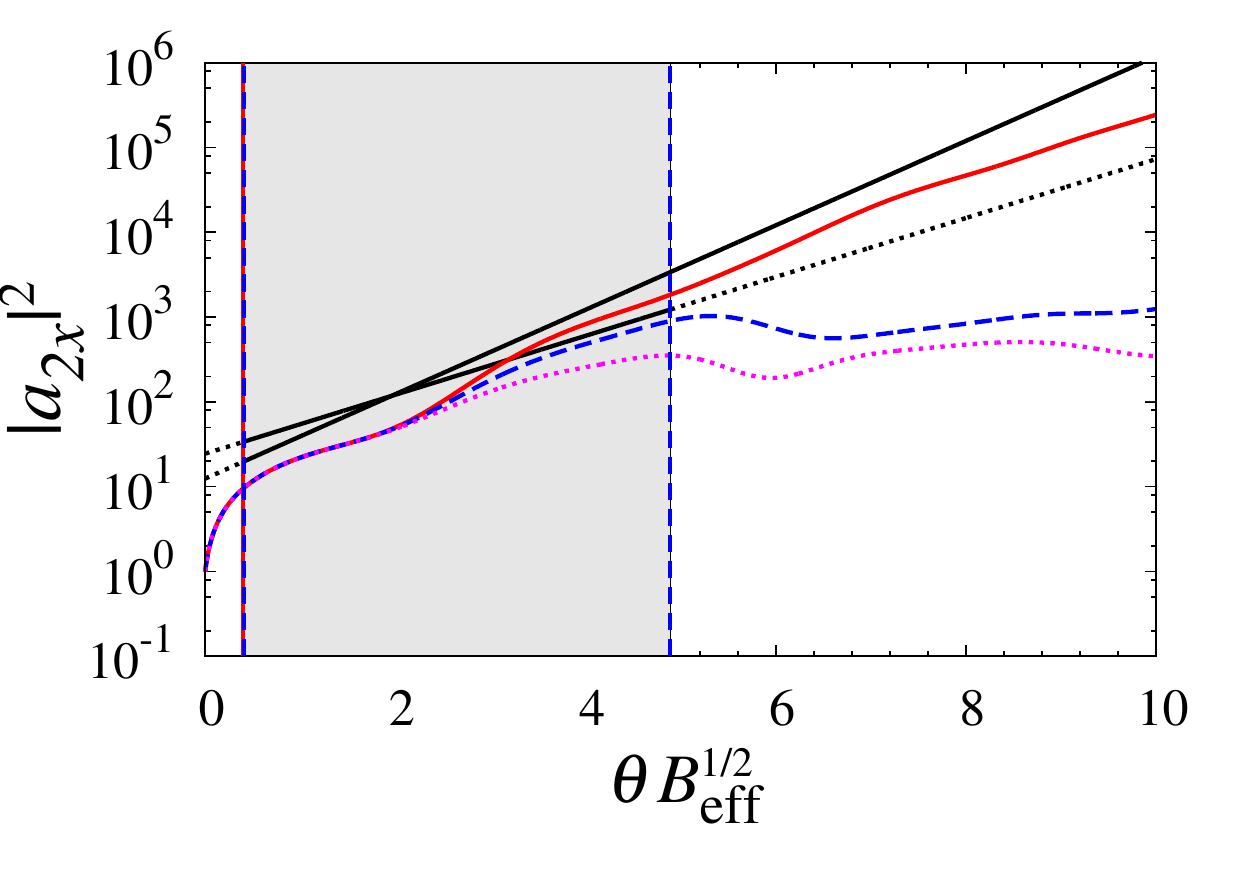}
	\PSfig{5.0cm}{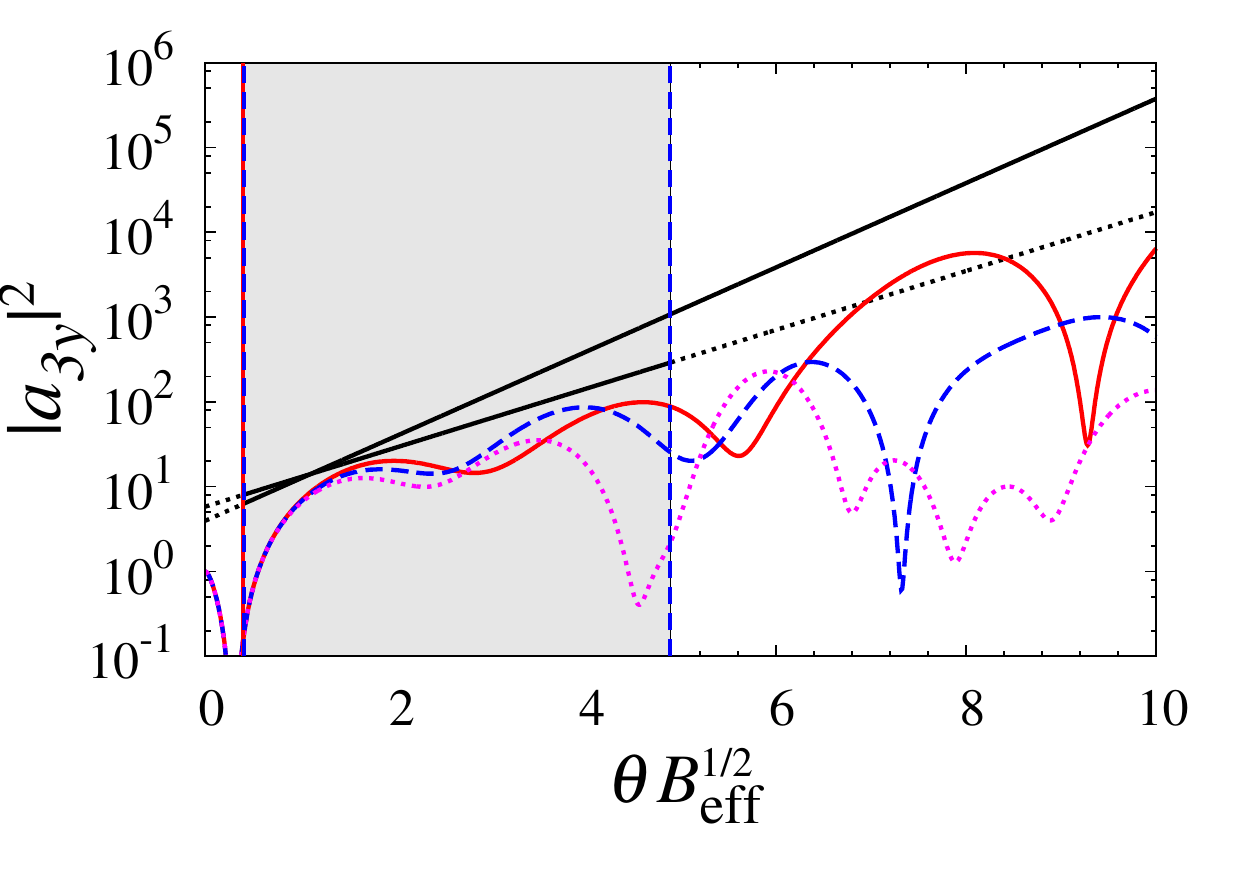}
	\PSfig{5.0cm}{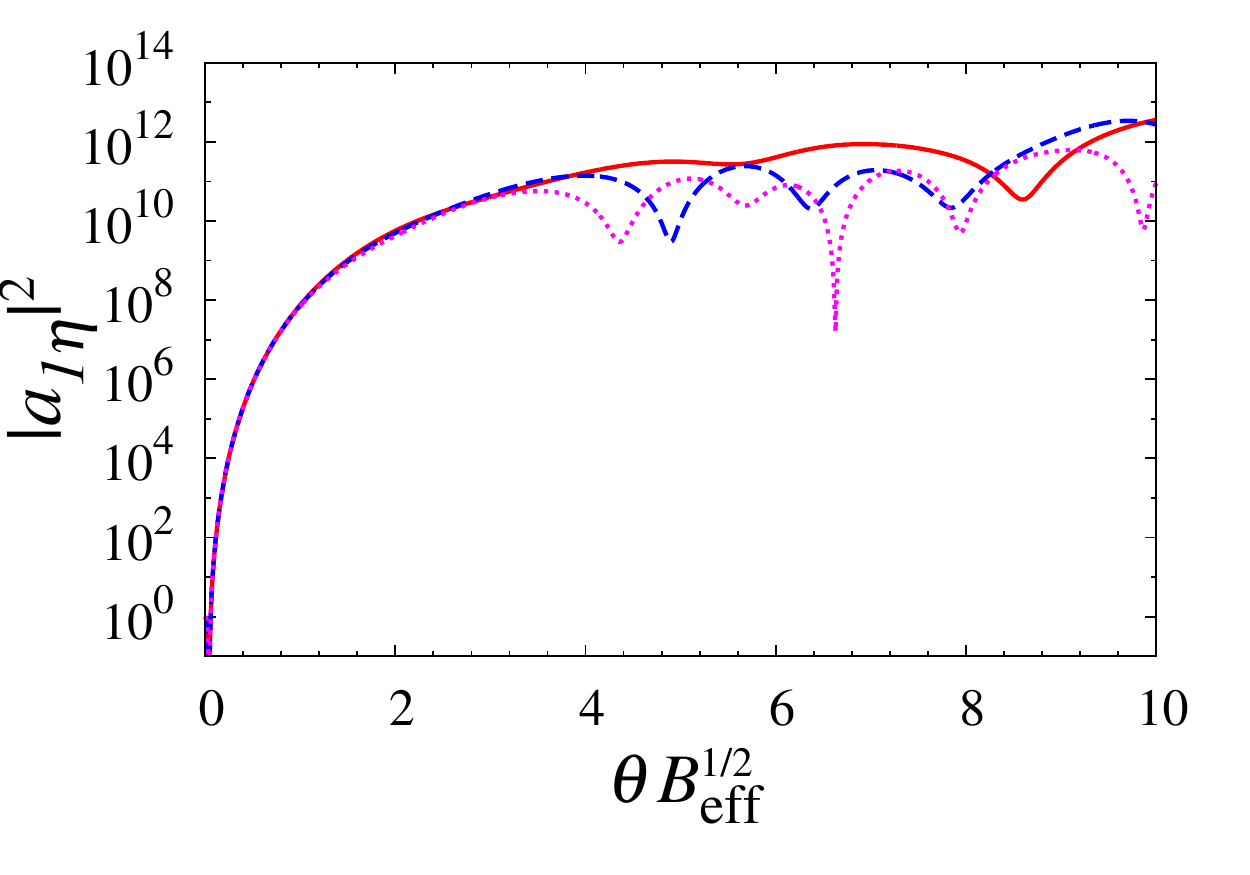}
	\PSfig{5.0cm}{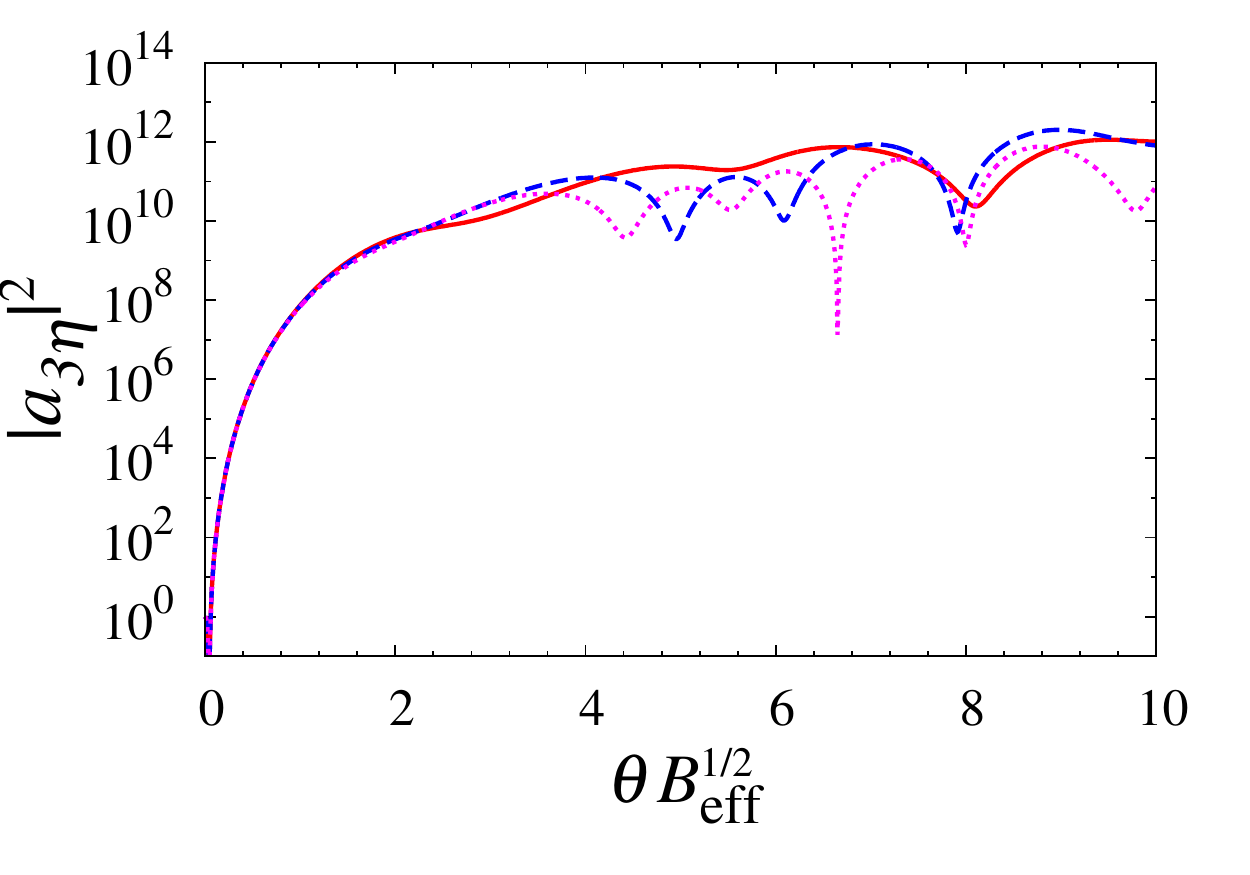}
	\PSfig{5.0cm}{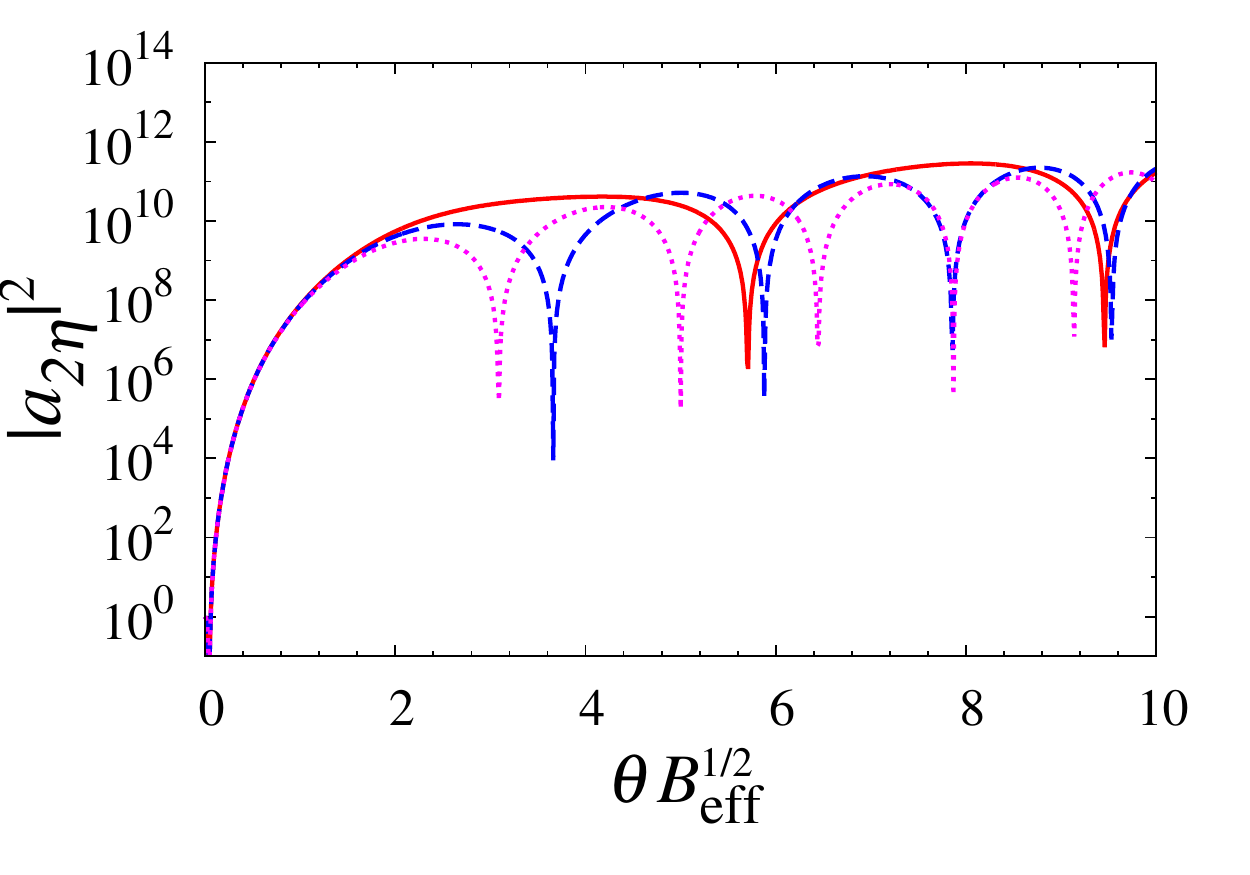}
	\caption{
	Squared fluctuation amplitudes in finite $p_T$ modes. 
	We show the time evolution of fluctuations in finite $p_T$ modes for $p^2_T = 0.1$, $0.5$ and $1.0$.
	In the upper panels, the growth rates of the function $e^{2\gamma\theta}$ are given by $\gamma = 0.57$, 0.43 and 0.31 for $p^2_T = 0.1$, 0.5 and 1.0, respectively.  
	In the middle panels, the growth rates are given by $\gamma = 0.57$ and 0.40 for $p^2_T = 0.1$ and 0.5, respectively.  
	Two transition times are indicated with vertical lines for each momentum mode.
	In the $\Omega^2_5$ sector which includes $a^1_x, a^2_y, a^3_x$,
	the first (second) transition time is given by 0.38 (26.2), 0.39 (5.22) and 0.41 (2.57) for $p^2_T = 0.1$, 0.5 and 1.0, respectively.
	In the $\Omega^2_4$ sector which includes $a^1_y, a^2_x, a^3_y$,
	the first (second) transition time is given by 0.39 (24.6) and 0.41 (4.89) for $p^2_T = 0.1$ and 0.5, respectively.
	Shaded areas show the time regime where exponential growth is expected for $p_\eta^2=0.5$.
	}
	\label{Fig:pT_trans}
\end{figure*}
%%%%%%%%%%%%%%%%%%%%%%%%%%%%%%%%%%%%%%%%%%%%%%%%%%%%%%%%%%%%%

\subsection{Numerical results}
\label{subsec:num}

We choose the initial conditions for fluctuations so that Gauss's law 
Eq.~\eqref{GL_conformal} is satisfied; which leads to
\begin{align}
	& i\tilp_i \dot{a}_i^a + \tiltheta_0^{-3} i\tilp_\eta \dot{a}_\eta^a
	- \frac{1}{3\tiltheta_0}
	\lp i\tilp_i a_i^a + \tiltheta_0^{-3} i\tilp_\eta a_\eta^a \rp \notag \\
	&+\epsilon^{abc} \delta^{b2} \lp \tilA(\theta_0) \dot{a}_x^c - \dot{\tilA}(\theta_0) a_x^c \rp \notag \\
	&+\epsilon^{abc} \delta^{b1} \lp \tilA(\theta_0) \dot{a}_y^c - \dot{\tilA}(\theta_0) a_y^c \rp
	=0 ,
\end{align}
where $\tilp_I=\tiltheta^{1/2}_0p_I$.
We can choose the initial transverse components as $a^a_i = 1$ and $\deltheta a^a_i = 0$ in the rescaled dimensionless unit.
The initial longitudinal components are determined by the Gauss's law.
The initial condition of the background field is the same as that shown in Sec.~\ref{subsec:bg};
$\theta_0 B_\text{eff}^{1/2} \simeq 0.01\times 4.68$,
$\tilA / \rtB \simeq 1/4.68$,
and $\deltheta\tilA / \rtB = 0$
in the rescaled dimensionless unit.

Figures~\ref{Fig:pL_trans} and~\ref{Fig:pT_trans} show the numerical results of squared fluctuation amplitudes obtained by solving the EOM given by Eq.~\eqref{linearEOM} for finite $p_\eta$ and finite $p_T$ modes, respectively.
For comparison, we show the exponential functions $e^{2\gamma\theta}$ by black solid lines in those modes where exponential growth is expected.
The comparison shows that there are several exponentially unstable modes.
The amplitudes in most unstable modes ($a_{A+}$ and $a_{B-}$ components in finite $p_\eta$ modes) obey the Lam\'e's equation with $\lambda=-1$ at later times, and have a growth rate $\gamma \simeq 0.66$.
The Lam\'{e}'s equation with $\lambda = -1$ describes the most unstable mode also in the nonexpanding geometry.
Therefore, we conclude that the parametric instability emerges also in the expanding geometry in almost the same way as in the nonexpanding geometry.
More details of the growth rates are discussed in Sec~\ref{Sec:growth}.

We also find that unstable modes grow exponentially after some times or in a limited time regime, and show oscillating behavior outside of the exponential regime.
The transitions between the oscillating behavior and the exponential growth occur at certain times marked by vertical lines in Figs.~\ref{Fig:pL_trans} and~\ref{Fig:pT_trans}.
For typical momenta, say $p^2_\eta=2.0$ and $p^2_T=0.5$, the exponential growth is observed in the shaded areas while fluctuation amplitudes just oscillate outside the areas.
The oscillatory behavior in the earliest stage is caused by the non-periodic evolution of the background field.
We discuss the evolution in the earliest stage in Sec~\ref{Sec:earliest}.
We also estimate the transition times in Subsecs.~\ref{Sec:transition} and~\ref{Sec:physical}.

%%%%%%%%%%%%%%%%%%%%%%%%%%%%%%%%%%%%%%%%%%%%%%%%%%%%%%%%%%%%%%
\subsection{Earliest stage}\label{Sec:earliest}
We discuss the behavior of fluctuations in the earliest stage characterized by $\theta\ll1$.
In this stage,
the effective momenta give rise to the most singular terms $k^2_\text{eff}\propto\theta^{-2}$,
which is much larger than the background field $\tilA \sim \theta^{1/2}$ as given in Eq.~\eqref{early_bg}.

The EOMs for finite $p_\eta$ modes in the earliest stage read
\begin{align}
	\ddot{a}^a_i + k^2_{\eta T}(\theta,p_\eta) a^a_i + \calO(\theta^{-2/3},\theta) &= 0, \\
	\calL^2_\eta a^a_\eta + M^2_{\eta}(\theta) a^a_\eta + \calO(\theta^{-2/3},\theta) &= 0.
\end{align}
Both of the effective momenta,
$k^2_{\eta T}$ and $k^2_{\eta\eta}=M^2_\eta$ given in Eqs.~\eqref{keT} and~\eqref{kee}, are proportional to $\theta^{-2}$.
The general solutions for the transverse components are given by $a^a_i = c \theta^{1/2 + 3ip_\eta/2} + \conj$ where $c$ is an arbitrary constant: 
One sees that the $p_\eta$ dependence of $k_{\eta T}^2$ causes the oscillating behavior of the solution.
The longitudinal components do not have $p_\eta$ dependence, and the general solutions are given by $a^a_\eta = c_1 \theta^{1/2} + c_2 \theta^{7/2}$.

For finite $p_T$ modes, we have
\begin{align}
	\deltheta^2 a^a_i + M_T^2(\theta) a^a_i + \calO(\theta) &= 0, \\
	\calL_\eta^2 a^a_\eta + M_\eta^2(\theta) a^a_\eta + \calO(\theta) &= 0.
\end{align}
In the earliest stage, $p_T$ dependence appears in $\calO(\theta)$ terms, as found in Eqs.~\eqref{kTT} and \eqref{kTe}.
Note that EOM of the transverse components has the same form as that of the background field, 
then the general solution is given by $a^a_i = c_1\theta^{1/2} + c_2\theta^{1/2}\log\theta$.
The solution for the longitudinal components grows in time with the same time dependence as the solution of finite $p_\eta$ modes.

From the above arguments, 
we find that fluctuation amplitudes in both finite $p_\eta$ and $p_T$ modes 
behave as $a \propto \theta^{1/2} +$ (higher order) in the earliest stage,
except for the logarithmic corrections in $\theta$.
The fluctuation amplitude in the $\tau$-$\eta$ coordinate is given by $\tau^{-1/3}a$ from Eqs.~\eqref{trans_Ai} and \eqref{trans_Aeta},
then they will not grow as a function of the proper time, i.e.
$\tau^{-1/3}a(\tau) \propto \tau^{-1/3} \theta(\tau)^{1/2} = \calO(\tau^0)$, 
if we ignore logarithmic corrections.
The numerical results are in good agreement with the above discussion.
In fact, no exponential growth is found in Figs.~\ref{Fig:pL_trans} and~\ref{Fig:pT_trans} at the earliest stage,

%%%%%%%%%%%%%%%%%%%%%%%%%%%%%%%%%%%%%%%%%%%%%%%%%%%%%%%%%%%%%%
\subsection{Growth rates}\label{Sec:growth}
We shall now discuss exponential instability of fluctuations.
In Figs.~\ref{Fig:pL_trans} and~\ref{Fig:pT_trans},
there are some exponentially growing modes, namely,
%\sout{the finite $p_\eta$ modes of}
$a_{A\pm}, a_{B-}$ and $a^3_\eta$ components 
of the the finite $p_\eta$ modes
and 
$a^a_x$ and $a^a_y$ 
components of the finite $p_T$ modes.
As we shall see below, 
the growth rates of them are nicely determined by utilizing the Floquet analysis  in the {\em nonexpanding} geometry which was worked out in the previous work~\cite{Tsutsui:2014rqa}:
The (time-dependent) effective momenta in the expanding geometry are mapped in the contour map of the instability bands for the nonexpanding geometry.
In the following analysis, 
we shall first recapitulate and utilize the results in the previous work for later convenience.

\begin{figure}[bth]
	\PSfig{7.0cm}{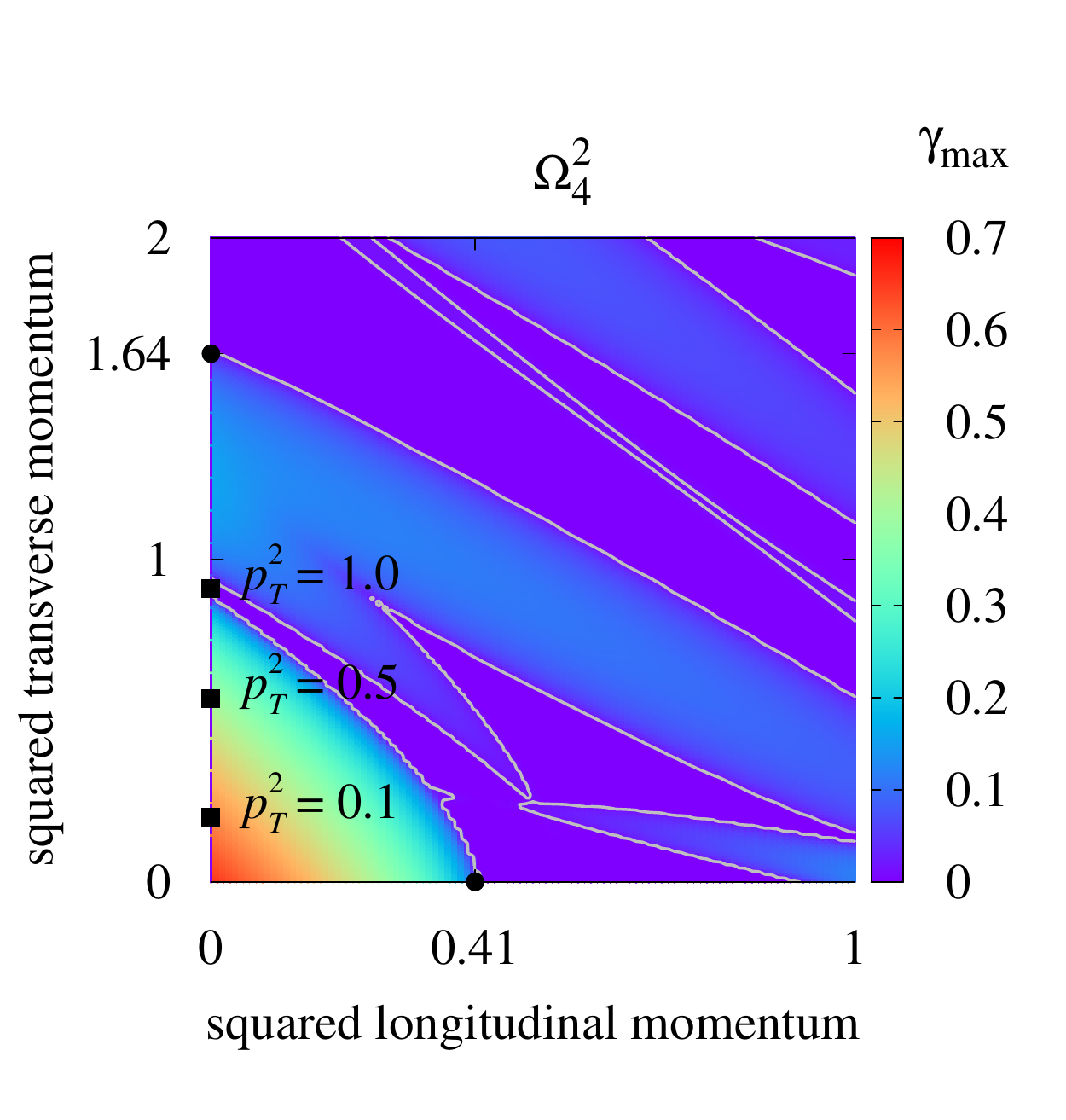}
	\PSfig{7.0cm}{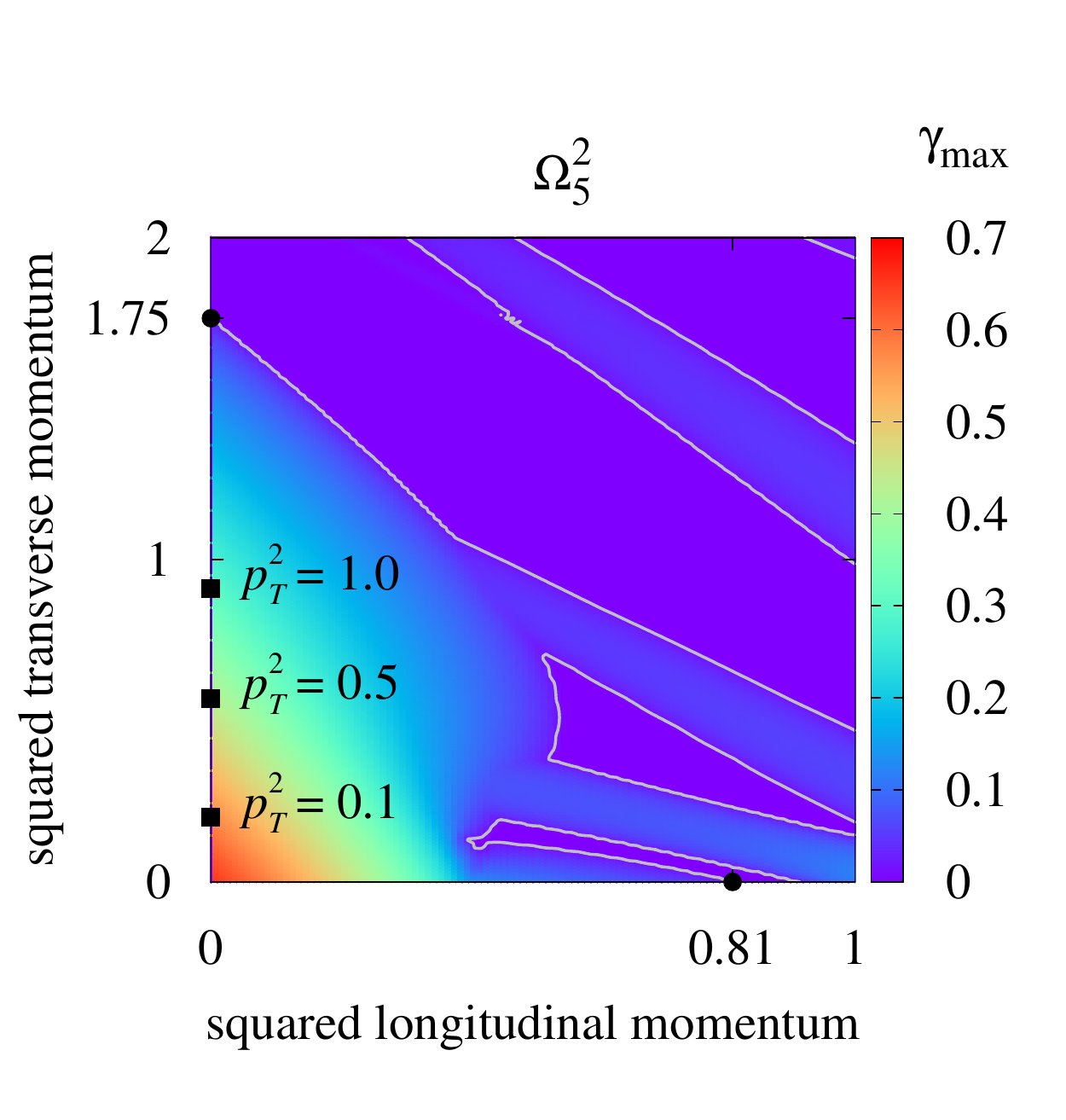}
	\caption{
		Maximal growth rates as functions of transverse and longitudinal momenta for $\Omega^2_4$ sector (top) and $\Omega^2_5$ sector (bottom) in 
		the nonexpanding geometry.
		The band boundaries denoted by gray solid lines are determined by $|\gamma|_\text{max}=0.01$.
		Black squares denote the minimum values of $k_{TT}$.
		In both sectors, $\min k_{TT} = 0.20$, 0.57 and 0.91 for $p^2_T = 0.1$, 0.5 and 1.0.
        On those points $\gamma = 0.57$ , 0.40 and 0.00 in $\Omega^2_4$ sector and 
        $\gamma = 0.57$, 0.43 and 0.31 in $\Omega^2_5$ sector.
	}
	\label{Fig:maps}
\end{figure}

In Fig.~\ref{Fig:maps}, we show the maximal growth rate in the nonexpanding geometry as a function of transverse and longitudinal momenta.
The contour maps are obtained for 
$\Omega^2_4$ and $\Omega^2_5$ sectors, separately, 
and the maximal rate among the several (4 and 5) growth rates
is shown in the figure.
Figure~\ref{Fig:maps} clearly shows the band structure of the parametric instability.
The boundaries of instability bands are depicted by gray solid lines.
The largest growth rate $\gamma_\mathrm{max}=0.66$ is found at $p=0$.
These results in the nonexpanding geometry enable us to estimate the growth rates of the numerical solutions in 
the expanding geometry by paying attention to the time dependence of the effective momenta.

Let us evaluate the growth rates in the finite $p_\eta$ modes
by using the growth rates in the nonexpanding geometry.
In the EOMs for finite $p_\eta$ modes summarized in Eqs.~\eqref{Eq:A+}-\eqref{Eq:C1eta},
we neglect terms with negative powers of $\theta$ and obtain the asymptotic EOMs as
\begin{align}
	\frac{d^2 a^a_I}{d\theta^2} + \lambda \cn^2(\theta+\Delta)\ a^a_I = \text{inhomogeneous terms},
	\label{Eq:peta_asy}
\end{align}
where $\lambda = \pm1, 2$ or $3$, and we have replaced the background field with its asymptotic form,
$\tilA \sim \cn(\theta+\Delta;1/\sqrt{2})$ shown in Eq.~\eqref{bg_asy}.
These equations are nothing but Lam\'{e}'s equation with inhomogeneous terms.
The solutions are unstable for $\lambda = -1$ and 2 
due to the parametric instability as we show in Fig.~\ref{Fig:lames}.
Their growth rates are given by those of the \textit{zero momentum} mode in the nonexpanding geometry, 
and are found to be $\gamma = 0.66$ and $\gamma = 0.23$ for $\lambda =-1$ and $\lambda =2$, respectively.
We summarize instability properties of finite $p_\eta$ modes in Table~\ref{tableL}.

For the finite $p_T$ modes, we need to be cautious about the non-monotonic
dependence of the effective momenta on $\theta$.
The EOMs of the finite $p_T$ modes are given in Eqs.~\eqref{Eq:D1x}$-$\eqref{Eq:G2eta}. 
The effective transverse momentum Eq.~\eqref{kTT} first decreases, takes a minimum at a finite positive value, and increases again with increasing $\theta$.
We expect that the maximum growth rate may be evaluated at the $k_{TT}$ minima.
The growth rates in the $\Omega^2_5$ ($\Omega^2_4$) sector at the minima of $k_{TT}$, marked by squares in Fig.~\ref{Fig:maps}, are $\gamma = 0.57$ (0.57), 0.43 (0.40) and 0.31 (0.00) for $p^2_T = 0.1$, 0.5 and 1.0, respectively.
The mode with $p^2_T = 1.0$ should be stable in the $\Omega^2_4$ sector.

%%%%%%%%%%%%%%%%%%%%%%%%%%%%%%%%%%%%%%%%%%%%%%%%%%%%%%%%%%%%%%%%%%%%
%%%%%%%%%%%%%%%%%%%%%%%%%%%%%%%%%%%%%%%%%%%%%%%%%%%%%%%%%%%%%%%%%%%%

We now examine the growth rates estimated from the Floquet analysis by comparing them with the numerical results in the expanding geometry.
In the upper panels of Fig.~\ref{Fig:pL_trans},
we compare the numerical results of finite $p_\eta$ modes with exponential functions $\exp(2\gamma \theta)$.
The numerical solutions are plotted for $p_\eta^2 = 1.0, 2.0$ and $3.0$.
The most unstable components are 
$a_{A+}=(a^1_x + a^2_y)/\sqrt{2}$ and $a_{B-} = (a^1_y - a^2_x)/\sqrt{2}$
whose growth rates are expected to be $\gamma=0.66$
from Lam\'{e}'s equation with $\lambda =-1$.
The next dominant unstable component is $a_\eta^3$
whose growth rate is expected to be $\gamma=0.23$ from
Lam\'{e}'s equation with $\lambda = 2$.
As expected, the growth rates for $a_{A+}$, $a_{B-}$ and $a_\eta^3$
in the expanding geometry approach towards the upper bound
determined by those in a nonexpanding geometry.
In principle, the $a_{A-}$ component can also show exponential growth from the inhomogeneous term containing $\theta^{-5/2} a^3_\eta$, but we do not see such instability.
More strict discussion on the growth rate based on the monodromy matrix is given in Appendix~\ref{App:growth}.

%%%%%%%%%%%%%%%%%%%%%%%%%%%%%%%%%%%%%%%%%%%%%%%%%
\begin{table}
	\centering 
	\begin{tabular}{|c|c||c|c|c|c|}
		\hline sector & component  & $\lambda$ & inhomo. term & instability & growth rate  \\ 
		\hline
		\hline $\Omega^2_5$ & $a_{A+}$ & -1 & inhomo & exponential & $\gamma=0.66$ \\ 
		\hline $\Omega^2_5$ & $a_{A-}$ & +1 & inhomo & (*)exponential & $\gamma=0.23$ \\ 
		\hline $\Omega^2_4$ & $a_{B+}$ & +3 & - & linear & - \\ 
		\hline $\Omega^2_4$ & $a_{B-}$ & -1 & - & exponential & $\gamma=0.66$ \\ 
		\hline $\Omega^2_5$ & $a^3_x$ & +1 & - & linear & - \\ 
		\hline $\Omega^2_4$ & $a^3_y$ & +1 & - & linear & - \\ 
		\hline $\Omega^2_5$ & $a^1_\eta$ & +1 & inhomo & linear & - \\ 
		\hline $\Omega^2_4$ & $a^2_\eta$ & +1 & inhomo & linear & - \\ 
		\hline $\Omega^2_5$ & $a^3_\eta$ & +2 & inhomo & exponential & $\gamma=0.23$ \\ 
		\hline 
	\end{tabular}
	\caption{The classification of the asymptotic behavior for $p_T=0$ modes.
		Here, $a_{A\pm} = (a^1_x \pm a^2_y)/\sqrt{2}$ and $a_{B\pm} = (a^1_y \pm a^2_x)/\sqrt{2}$.
		(*)For $p_\eta=0$, the inhomogeneous term vanishes and show only linear divergence as ordinal instability of Lam\'{e}'s 
%\com{
equation with $\lambda = 1$.
%} 
	}
	\label{tableL}
\end{table}
%%%%%%%%%%%%%%%%%%%%%%%%%%%%%%%%%%%%%%%%%%%%%%%%%
%
%	

%\sout{In the upper panels of Fig.~\ref{Fig:pL_trans}, 
%we show exponential functions whose growth rates are given in Table~\ref{tableL} by black lines.}

Let us turn to the finite $p_T$ modes.
In Fig.~\ref{Fig:pT_trans},
we compare the exponential functions $\exp(2\gamma \theta)$ with numerical solutions.
Here, $a^1_y$, $a^2_x$ and $a^3_y$ belong to $\Omega^2_4$ sector 
and $a^1_x$, $a^2_y$ and $a^3_x$ belong to $\Omega^2_5$ sector, respectively.
The values of $\gamma$  estimated at the minima of $k_{TT}$ give the upper bound of the growth rates of these unstable modes.
The growth rates get smaller at later times,
where the effective momenta become large.
This behavior is in accordance with the fact that there is no significant instability band
in the high transverse momentum region.

\subsection{transition times}\label{Sec:transition}
We now discuss the temporal regime where the exponential growth is expected.
The main difference between nonexpanding and expanding geometries
comes from the time dependence of the effective momenta 
$k_\text{eff}^2=k^2_{\eta T}, k^2_{\eta\eta}, k^2_{TT}$ or $k^2_{T\eta}$,
defined in Eqs.~ \eqref{keT},~\eqref{kee},~\eqref{kTT} and~\eqref{kTe}. 
They control the transition between oscillating behavior and the exponential growth;
When $k^2_\text{eff}$ is in the instability bands of the corresponding nonexpanding case, we can expect exponential growth.
Shaded areas in Figs.~\ref{Fig:pL_trans} and \ref{Fig:pT_trans} show the temporal regime, where exponential growth is expected.

%%%%%%%%%%%%%%%%%%%%%%%%%%%%%%%%%%%%%%%%%%%%%%%%%%%%%%%%%%%%%%%%%%%%
\begin{figure}[bth]
	\PSfig{8.0cm}{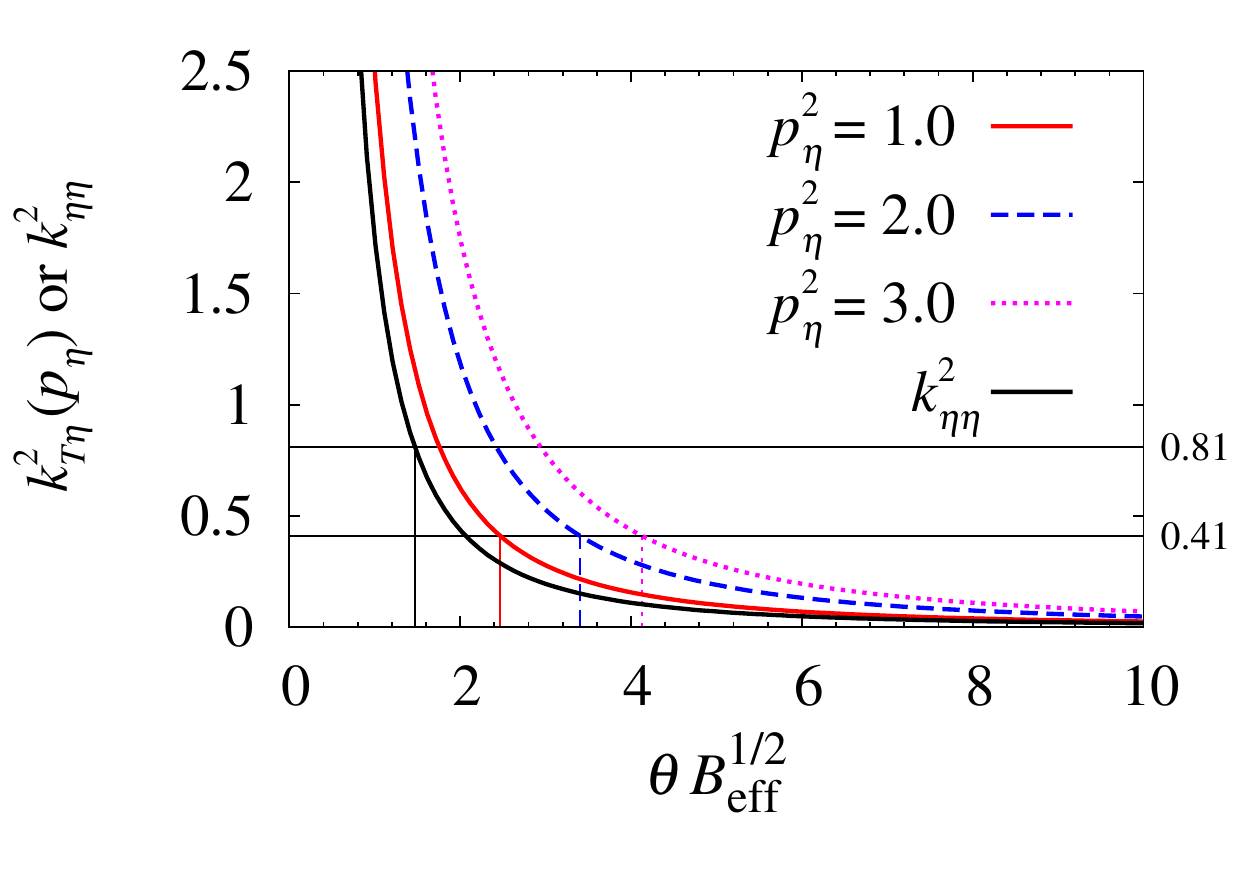}\\
	\PSfig{8.0cm}{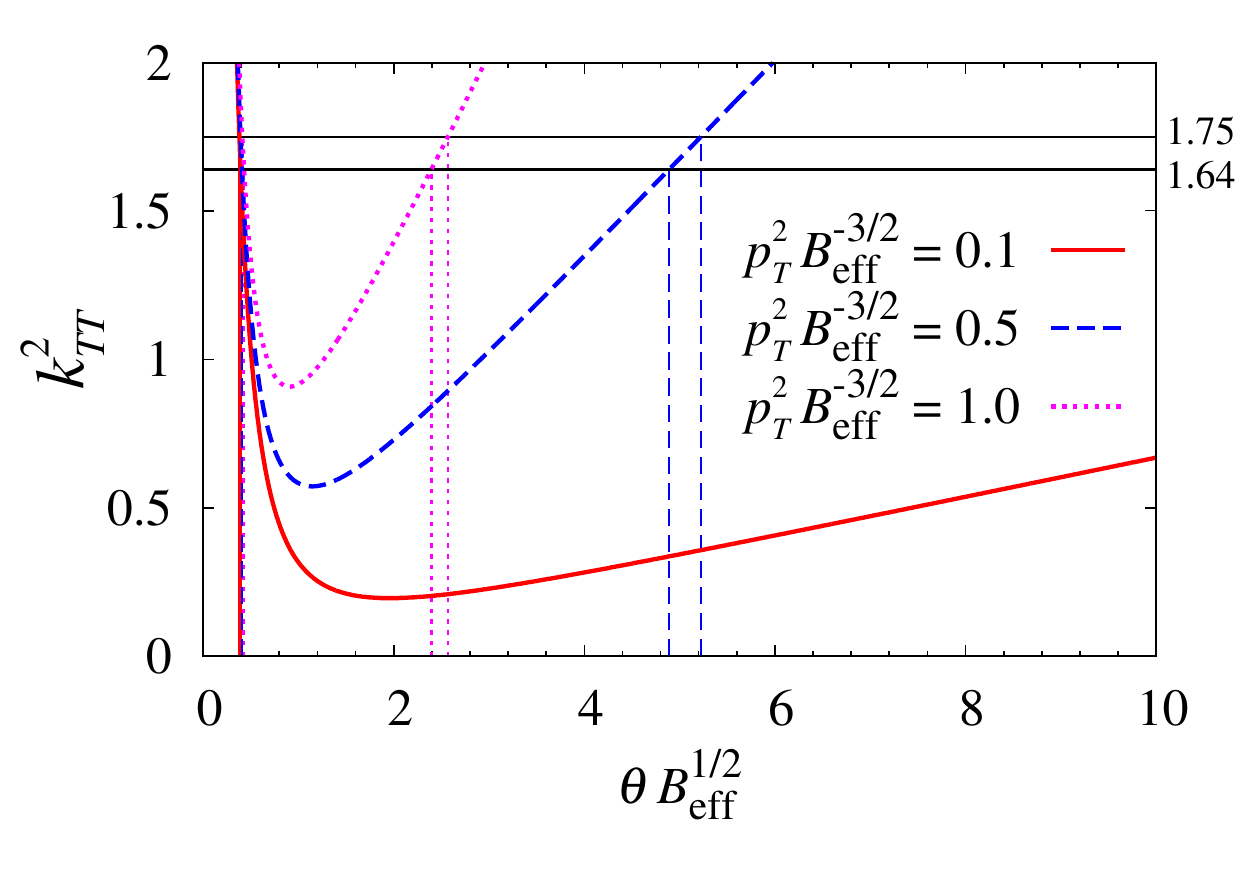}
	\caption{
	%The 
	Time dependence of the effective momenta.
	The horizontal lines show the boundaries of the instability bands.
	(Upper panel)
	Effective momenta in finite $p_\eta$ modes.
	The boundaries are given by $k^2_{T\eta} (\text{or } k^2_{\eta\eta}) = p^2_{b\eta} = 0.81$ and $k^2_{T\eta} (\text{or } k^2_{\eta\eta}) = 0.41$. 
	(Lower panel)
	Effective momenta of finite $p_T$ modes.
	The boundaries are given by $k^2_{TT} = p^2_{bT} = 1.75$ and $k^2_{TT} = 1.64$. 
	In both panels, vertical lines show the transition times.
	}
	\label{Fig:effmom}
\end{figure}
%%%%%%%%%%%%%%%%%%%%%%%%%%%%%%%%%%%%%%%%%%%%%%%%%%%%%%%%%%%%%%%%%%%%

Figure~\ref{Fig:effmom} shows the time dependence 
of the effective momenta $k^2_{\eta T}$, $k^2_{\eta \eta}$ and $k^2_{TT}$.
The horizontal lines correspond to the boundaries of the instability bands.
For finite $p_\eta$ modes,
the effective momenta~\eqref{keT} and~\eqref{kee} decrease in time,
then the infrared band structure of the nonexpanding geometry
is relevant and responsible for the emergence of instabilities.
Two horizontal lines in the top panel of Fig.~\ref{Fig:effmom} represent the longitudinal momenta on the boundaries of the instability band 
around the zero momentum region;
0.41 for $\Omega^2_4$ sector and 0.81 for $\Omega^2_5$ sector.
Using the momentum on the boundary point $p_{b\eta}$, 
we define the transition time $\theta_\text{tr}$ by 
\begin{align}
p^2_{b\eta} &= k^2_{\eta T}(\theta_\text{tr}) = M^2_T(\theta_\text{tr}) + 9p^2_\eta/4\theta_\text{tr}^2, \\
p^2_{b\eta} &= k^2_{\eta\eta}(\theta_\text{tr}) = M^2_\eta(\theta_\text{tr}),
\end{align}
for transverse and longitudinal components, respectively.
Note that the latter has no $p_\eta$ dependence.
The transition times at $p^2_\eta = 1.0$ are $\theta_\text{tr} = 2.46$ (1.46) 
for $a_{A+}$ and $a_{B-}$ ($a_\eta^3$) components, respectively.

Finite $p_T$ modes can also show instability.
Two horizontal lines in the lower panel of Fig.~\ref{Fig:effmom} 
represent the transverse momenta on the band boundaries;
 1.64 for $\Omega^2_4$ sector and 1.75 for $\Omega^2_5$ sector.
All unstable components are transverse $a^a_i$,
so the transition time $\theta_\text{tr}$ is defined by
\begin{align}
p^2_{bT} = k^2_{TT}(\theta_\text{tr}) = M^2_T(\theta_\text{tr}) + \frac{2}{3}\theta_\text{tr} p_T^2 .
\end{align}
Since effective momenta are not monotonic in time,
there are two transition times, between which exponential growth is expected.
For instance, the first transition time at $p^2_T = 0.5$ is $\theta_\text{tr} = 0.41$ (0.39) 
for the $\Omega^2_4$ ($\Omega^2_5$) sector, 
which is the starting time of the exponential growth.
The second transition time is $\theta_\text{tr} = 4.9$ (5.2) 
for the $\Omega^2_4$ ($\Omega^2_5$) sector,
which is the finishing time of the exponential growth.

In Figs.~\ref{Fig:pL_trans} and~\ref{Fig:pT_trans},
the transition times are denoted by vertical lines.
Our estimations well agree with the transitional behavior
of the numerical results.

%%%%%%%%%%%%%%%%%%%%%%%%%%%%%%%%%%%%%%%%%%%%%%%%%%%%%%%%%%%%%%
\subsection{Physical time scale}\label{Sec:physical}
Finally, we should comment on the physical scale of the transition times.
As we have seen in Sec.~\ref{sec:EOM},
all quantities are scaled by the strength of the background magnetic field 
$B_\text{eff}$ in our setup.
By using the relation between the conformal and proper times in Eq.~\eqref{Eq:timescale} and the empirical value of the saturation scale,
$Q_\text{s} \simeq 1$ GeV for RHIC or $Q_\text{s} \simeq 2$ GeV for LHC,
the transition time for the fastest growth modes ($a_{A+}$ and $a_{B-}$) is estimated as
\begin{align}
\tau_t=&\frac23\, \left(\theta_t \rtB\right)^{3/2}\,
\frac{1}{Q_\mathrm{s}}
\nonumber\\
\simeq&
\begin{cases}
 0.53 \,(1.12)~\mathrm{fm}/c & \text{for RHIC,} \\
 0.26 \,(0.56)~\mathrm{fm}/c & \text{for LHC,}
\end{cases}
\label{phys_time}
\end{align}
for $p_\eta^2=1.0 \,(3.0)$.

These fastest growth modes can emerge in the dynamics of relativistic heavy-ion collisions.
In heavy-ion experiments,
the rapidity gaps between the projectile and target are about $\Delta Y=10.7$ in RHIC and $17.4$ in LHC.
The rapidity gap gives the lower bound of the longitudinal momentum of fluctuations;
In order to observe half wave length of the fluctuation, $\Lambda_\eta \Delta Y \sim \pi$ is required.
For RHIC and LHC,
we find $\Lambda_\eta^2 \sim 0.08$ and $\Lambda_\eta^2 \sim 0.03$, respectively.
The typical momenta of the fastest growth modes, say $p_\eta^2 = 1.0 (3.0)$, are sufficiently larger than these lower bounds.

These results suggest that the parametric instability, 
especially the instability in the fastest growing modes,
might be relevant in the dynamics of CYM field in an expanding geometry.

%%%%%%%%%%%%
%%%%%%%%%%%%%%%%%%%%%%%%%%%%%%%%%%%%%%%%%%%%%%%%%%%%%%%%%%%%%%
\section{Conclusions}\label{sec:Conc}
We have studied the instability of CYM field in an expanding geometry under the longitudinal color magnetic background which is homogeneous, boost-invariant and time-dependent.
We have introduced the conformal variables
and map the expanding problem approximately into the nonexpanding problem.
The background gauge field at later times is described by the elliptic function in the conformal coordinate.
The main difference between nonexpanding and expanding geometries is the time dependence of the effective momenta.
We have performed a linear stability analysis of fluctuations on top of the oscillating background field.

Exponential growth has been found in some modes in the expanding geometry. We have elucidated that the fluctuations in the expanding geometry at later times obey the same EOM and have the same maximal growth rate as those in the nonexpanding geometry. Thus, we conclude that the expanding system under the time-dependent background field shows parametric instability as observed in the nonexpanding geometry.

We have also found that
the way of growth is qualitatively different between finite $p_\eta$ and finite $p_T$ modes.
We have made semi-analytic analyses on the instability in the expanding geometry by using the time-dependent effective momenta and the Floquet theory in a nonexpanding geometry.
Since the longitudinal effective momentum decreases monotonically, 
unstable fluctuations with finite $p_\eta$ grow exponentially after the transition from the oscillating behavior in the earliest stage.
On the other hand, 
since the transverse effective momentum is not monotonic in time, exponential growth in finite $p_T$ modes is limited in a particular temporal regime.
In this regime, effective momenta are small and located within the instability band, then fluctuations can grow exponentially.
In accordance with the band structure found in the nonexpanding geometry,
only the modes with small $p_T$ show significant instability.
Finally, we have estimated the physical scale of the transition times.
For finite $p_\eta$ modes,
the typical unstable modes start to grow at $\tau\simeq0.26 (0.53)$ fm$/c$ at LHC (RHIC) energy.
These results suggest that the parametric instability could be relevant to the gluodynamics in heavy-ion collisions.

One of the notable points is that the parametric instability emerges at early times in some classes of field theories as discussed, for example, in cosmic inflation \cite{Amin:2014eta}.
The appearance of the parametric instability seems to be a universal phenomenon in field theories including the Yang-Mills theory.
While we have established that the parametric instability emerges under the color magnetic background fields in the homogeneous system, there are still unsolved and interesting questions.
One is associated with the robustness of the parametric instability.
It is nontrivial whether this instability is relevant or not to inhomogeneous systems, in particular glasma evolution from a more realistic initial condition.
It is also important to investigate how the parametric instability affects the particle production where the nonlinear interaction of the gluon fluctuations must be taken into account.
Investigations of these problems are beyond the scope of the present
work and left as future problems.

\section*{ACKNOWLEDGMENTS}
S.T. is supported by the Grant-in-Aid for JSPS fellows (No.26-3462).
This work was supported in part by 
the Grants-in-Aid for Scientific Research from JSPS 
(Nos. 23340067, %((B) PI:T.Kunihiro),
15K05079, %((C) PI:A.Ohnishi w/ Y.Nara)),
and
15H03663),  %((B) PI:A.Nakamura w/ T.Hatsuda, T.Kunihiro, A.Ohnishi),
the Grants-in-Aid for Scientific Research on Innovative Areas from MEXT
(No. 2404: 24105001, 24105008),
and by the Yukawa International Program for Quark-Hadron Sciences.
),

\appendix
\section{The linearized EOM of fluctuations in an expanding geometry}\label{App:detail}
In this appendix, 
we show the 
%\sout{specific}\com{explicit}
explicit form of the linearized EOM of fluctuations in the color magnetic background Eq.~\eqref{bgconf}.
The symbolic form of the EOM is given by Eq.~\eqref{linearEOM}.
Without loss of generality,
the coefficient matrix $\Omega^2$ is decomposed into two independent sectors $\Omega_4^2$ and $\Omega_5^2$ due to the rotational symmetry 
%\sout{of}\com{in the}
in the transverse direction. 
In the momentum representation, the explicit forms of them are given by
\begin{widetext}
\begin{align}
	([\Omega_4^2]_{\alpha\beta}) &= 
	\begin{pmatrix}
		%(1y)-ROW 
		\tilp_x^2 + \tiltheta^{-3}\tilp_{\eta}^2 + \tilA^2 & 2\tilA^2 & 0 & -2i\tilA \tilp_x \\
		%(2x)-ROW 
		2\tilA^2 & \tiltheta^{-3}\tilp_{\eta}^2 + \tilA^2 & -\tiltheta^{-3}\tilp_\eta \tilp_x &  -i\tilA \tilp_x \\
		%(2z)-ROW 
		0 & -\tiltheta^{-3}\tilp_\eta \tilp_x & \tiltheta^{-3}\lp \tilp_x^2 + \tilA^2 \rp & -\tiltheta^{-3}i\tilA \tilp_\eta \\
		%(3y)-ROW 
		2i\tilA \tilp_x & i\tilA \tilp_x & \tiltheta^{-3}i\tilA \tilp_\eta & \tilp_x^2 + \tiltheta^{-3}\tilp_{\eta}^2 + \tilA^2  
	\end{pmatrix},
	\label{o4}
	\\
	([\Omega_5^2]_{AB}) &= 
	\begin{pmatrix}
		%(1x)-ROW 
		\tiltheta^{-3}\tilp_{\eta}^2  & -\tiltheta^{-3}\tilp_\eta \tilp_x & -\tilA^2 & 0 & \tiltheta^{-3}i\tilA \tilp_\eta  \\
		%(1z)-ROW 
		-\tiltheta^{-3}\tilp_\eta \tilp_x & \tiltheta^{-3} \lp \tilp_x^2 + \tilA^2 \rp & 0 & \tiltheta^{-3}i\tilA \tilp_\eta  & -\tiltheta^{-3}2i\tilA \tilp_x\\
		%(2y)-ROW 
		-\tilA^2 & 0 & \tilp_x^2 + \tiltheta^{-3}\tilp_{\eta}^2 & -i\tilA \tilp_x & -\tiltheta^{-3}i\tilA \tilp_\eta  \\
		%(3x)-ROW 
		0 & -\tiltheta^{-3}i\tilA \tilp_\eta & i\tilA \tilp_x &  \tiltheta^{-3}\tilp_{\eta}^2 + \tilA^2 & -\tiltheta^{-3}\tilp_\eta \tilp_x \\
		%(3z)-ROW 
		-\tiltheta^{-3}i\tilA \tilp_\eta  & \tiltheta^{-3}2i\tilA \tilp_x & \tiltheta^{-3}i\tilA \tilp_\eta  & -\tiltheta^{-3}\tilp_\eta \tilp_x & \tiltheta^{-3} \lp \tilp_x^2 + 2\tilA^2 \rp 
	\end{pmatrix},
	\label{o5}
\end{align}
\end{widetext}
where $\tilp_I=\tiltheta^{1/2}p_I$ and we use following notation: 
$\alpha, \beta, \dots = (1y,2x,2\eta,3y)$ and
$A, B, \dots = (1x,1\eta,2y,3x,3\eta)$.

For the sake of the stability analysis performed in Sec.~\ref{sec:Stability},
we consider two specific limits $p_\eta\neq0, p_T=0$ (finite $p_\eta$ modes) and $p_\eta=0, p_T\neq0$ (finite $p_T$ modes).
In these limits, the coefficient matrices \eqref{o4} and \eqref{o5}
are further decomposed to lower rank matrices:
\begin{align}
	\Omega^2_4 &= 
	\begin{cases}
		\diag(\Omega^2_B,\Omega^{*2}_C) \quad
	(\text{finite $p_\eta$ modes}) \\
		\diag(\Omega^2_E,\Omega^2_G) \quad
	(\text{finite $p_T$ modes})
	\end{cases},
	\label{o4_reduction}
	\\
	\Omega^2_5 &= 
	\begin{cases}
		\diag(\Omega^2_A,\Omega^2_C) \quad
	(\text{finite $p_\eta$ modes}) \\
		\diag(\Omega^2_D,\Omega^2_F) \quad
	(\text{finite $p_T$ modes})
	\end{cases}	.
	\label{o5_reduction}
\end{align}

In the case of finite $p_\eta$ modes ($p_\eta\neq0, p_T=0$),
the EOM is decomposed into four independent sectors $A, B, C$ and $C^*$.
The EOM of the $A$-sector is given by
\begin{align}
	\ddot{a}_{A+} + k^2_{\eta T} a_{A+} - \tilA^2 a_{A+} = 0, 
\label{Eq:A+}\\
	\ddot{a}_{A-} + k^2_{\eta T} a_{A-} + \tilA^2 a_{A-} + \sqrt{2}i\tiltheta^{-5/2}p_\eta\tilA a^3_\eta = 0, 
\label{Eq:A-}\\
	\calL^2_\eta a^3_\eta + M^2_\eta a^3_\eta
	+ 2\tilA^2 a^3_\eta - \sqrt{2}i\tiltheta^{1/2}p_\eta\tilA a_{A-} = 0,
\label{Eq:A3eta} \\
\calL^2_\eta = \frac{d^2}{d\theta^2} - \frac{2}{\tiltheta}\frac{d}{d\theta}, 
\end{align}
where $a_{A\pm} = (a^1_x \pm a^2_y)/\sqrt{2}$, $k^2_{\eta T} = M^2_T + 9p^2_\eta/4\theta^2$ and $M^2_T = M^2_x = M^2_y = 1/4\theta^2$.
Dots denote derivatives with respect to conformal time $\theta$.
The EOM of the $B$-sector reads
\begin{align}
	\ddot{a}_{B+} + k^2_{\eta T} a_{B+} + 3\tilA^2 a_{B+} = 0,
\label{Eq:B+}\\
	\ddot{a}_{B-} + k^2_{\eta T} a_{B-} - \tilA^2 a_{B-} = 0,
\label{Eq:B-}
\end{align}
where $a_{B\pm} = (a^1_y \pm a^2_x)/\sqrt{2}$.
The EOMs of the $C$ and $C^*$-sectors have the same form:
For the $C$-sector,
\begin{align}
	\ddot{a}^3_x + k^2_{\eta T} a^3_x + \tilA^2 a^3_x - i\tiltheta^{-5/2}p_\eta\tilA a^1_\eta = 0,
\label{Eq:C3x}\\
	\calL^2_\eta a^1_\eta + M^2_\eta a^1_\eta
	+ \tilA^2 a^1_\eta + i\tiltheta^{1/2}p_\eta\tilA a^3_x = 0,
\label{Eq:C1eta}
\end{align}
and for the $C^*$-sector,
\begin{align}
	\ddot{a}^3_y + k^2_{\eta T} a^3_y + \tilA^2 a^3_y + i\tiltheta^{-5/2}p_\eta\tilA a^2_\eta = 0,
	\label{Eq:Cs3y}\\
	\calL^2_\eta a^2_\eta + M^2_\eta a^2_\eta
	+ \tilA^2 a^2_\eta - i\tiltheta^{1/2}p_\eta\tilA a^3_y = 0,
	\label{Eq:Cs2eta}
\end{align}
respectively.
We note that $M^2_\eta = 7/4\theta^2$.

In the case of finite $p_T$ modes ($p_\eta=0, p_T\neq0$),
the EOM is decomposed into four independent sectors $D, E, F$ and $G$.
For of the $D$ and $E$-sector, we get
\begin{align}
	\ddot{a}^1_x + M^2_T a^1_x - \tilA^2 a^2_y = 0,
\label{Eq:D1x}\\
	\ddot{a}^2_x + k^2_{TT} a^2_y - \tilA^2 a^1_x - i\tiltheta^{1/2}p_x\tilA a^3_x = 0,
\label{Eq:D2x}\\
	\ddot{a}^3_x + M^2_T a^3_x + \tilA^2 a^3_x + i\tiltheta^{1/2}p_x\tilA a^2_y = 0,
\label{Eq:D3x}
\end{align}
and 
\begin{align}
	\ddot{a}^1_y + k^2_{TT} a^1_y + \tilA^2 a^1_y + 2\tilA^2 a^2_x - 2i\tiltheta^{1/2}p_x\tilA a^3_y = 0,
\label{Eq:E1y}\\
	\ddot{a}^2_y + M^2_T + \tilA^2 a^2_x + 2\tilA^2 a^1_y - i\tiltheta^{1/2}p_x\tilA a^3_y = 0,
\label{Eq:E2y}\\
	\ddot{a}^3_y + k^2_{TT} a^3_y + \tilA^2 a^3_y + i\tiltheta^{1/2}p_x\tilA (2a^1_y+a^2_x) = 0,
\label{Eq:E3y}
\end{align}
respectively.
The EOM of the $F$ and $G$-sector is given by
\begin{align}
	\calL^2_\eta a^1_\eta + k^2_{T\eta} a^1_\eta + \tilA^2 a^1_\eta - 2i\tiltheta^{1/2}p_x\tilA a^3_\eta = 0,
\label{Eq:F1eta}\\
	\calL^2_\eta a^3_\eta + k^2_{T\eta} a^3_\eta + 2\tilA^2 a^3_\eta + 2i\tiltheta^{1/2}p_x\tilA a^1_\eta = 0,
\label{Eq:F3eta}
\end{align}
and
\begin{align}
	\calL^2_\eta a^2_\eta + k^2_{T\eta} a^2_\eta + \tilA^2 a^2_\eta = 0,
\label{Eq:G2eta}
\end{align}
respectively.
Here we use the notion of the effective momenta, $k^2_{TT} = M^2_T + 2\theta p^2_x/3$
and $k^2_{T\eta} = M^2_\eta + 2\theta p^2_x/3$.

In the following subsections,
we give the solutions of the EOM in the early stage $\theta\ll 1$ and the late stage $\theta\gg 1$.

\subsection{$\theta\ll 1$}
Assuming $\theta\ll 1$,
the dominant term with respect to $\theta$ is of the order $\calO(\theta^{-2})$.
Recalling $\tilA \sim \calO(\theta^{1/2})$,
other terms depending on $\theta$ explicitly are of higher order in $\theta$.
Collecting leading order terms,
we get
\begin{align}
	\ddot{a}_{A+} + k^2_{\eta T} a_{A+} + \calO(\theta) = 0, \\
	\ddot{a}_{A-} + k^2_{\eta T} a_{A-} + \calO(\theta^{-3/2}) = 0, \\
	\calL^2_\eta a^3_\eta + M^2_\eta a^3_\eta + \calO(\theta) = 0,
\end{align}
for the $A$-sector and
\begin{align}
	\ddot{a}_{B+} + k^2_{\eta T} a_{B+} + \calO(\theta) = 0, \\
	\ddot{a}_{B-} + k^2_{\eta T} a_{B-} + \calO(\theta) = 0,
\end{align}
for the $B$-sector.
The EOMs of the $C$ and $C^*$-sectors are given by
\begin{align}
	\ddot{a}^3_x + k^2_{\eta T} a^3_x + \calO(\theta^{-3/2}) = 0, \\
	\calL^2_\eta a^1_\eta + M^2_\eta a^1_\eta + \calO(\theta) = 0,
\end{align}
and
\begin{align}
	\ddot{a}^3_y + k^2_{\eta T} a^3_y + \calO(\theta^{-3/2}) = 0, \\
	\calL^2_\eta a^2_\eta + M^2_\eta a^2_\eta + \calO(\theta) = 0,
\end{align}
respectively.
In summary,
when $p_\eta\neq0, p_T=0$,
all transverse components satisfy
\begin{align}
	\ddot{a}^a_i + k^2_{\eta T} a^a_i = 0,
\end{align}
while all the longitudinal components obey
\begin{align}
\calL^2_\eta a^a_\eta + M^2_\eta a^a_\eta = 0.
\end{align}
Thus their general solutions are given by
$a^a_i = c_1 \theta^{1/2 + 3ip_\eta/2} + \conj$
and 
$a^a_\eta = c_1 \theta^{1/2} + c_2 \theta^{7/2}$
, where ${c_i}$ are arbitrary constants.

In the case of finite $p_T$ modes,
the $p_T$ dependence vanishes for $k^2_{TT} \rightarrow M^2_T$ and $k^2_{T\eta} \rightarrow M^2_\eta$ as $\theta \rightarrow 0$.
This leads to
\begin{align}
	\ddot{a}^1_x + M^2_T a^1_x + \calO(\theta) = 0, \\
	\ddot{a}^2_y + M^2_T a^2_y + \calO(\theta) = 0, \\
	\ddot{a}^3_x + M^2_T a^3_x + \calO(\theta) = 0,
\end{align}
for the $D$-sector and
\begin{align}
	\ddot{a}^1_y + M^2_T a^1_y + \calO(\theta) = 0, \\
	\ddot{a}^2_x + M^2_T a^2_x + \calO(\theta) = 0, \\
	\ddot{a}^3_y + M^2_T a^3_y + \calO(\theta) = 0,
\end{align}
for the $E$-sector.
We also obtain
\begin{align}
	\calL^2_\eta a^1_\eta + M^2_\eta a^1_\eta + \calO(\theta) = 0, \\
	\calL^2_\eta a^3_\eta + M^2_\eta a^3_\eta + \calO(\theta) = 0,
\end{align}
for the $F$-sector and
\begin{align}
	\calL^2_\eta a^2_\eta + M^2_\eta a^2_\eta + \calO(\theta) = 0,
\end{align}
for the $G$-sector.
In summary,
when $p_\eta=0, p_T\neq0$,
all transverse components 
%\sout{follow}\com{obey} 
obey
\begin{align}
	\ddot{a}^a_i + M^2_T a^a_i = 0,
\end{align}
and the general solution is given by $a^a_i = c_1 \theta^{1/2} + c_2 \theta^{1/2} \log\theta$.
The solutions of longitudinal fluctuations are equivalent to the case of $p_\eta\neq0, p_T=0$.

\subsection{$\theta\gg 1$}
Assuming 
%\com{that} 
that $\theta\gg 1$,
terms involving $\theta$ with a negative exponent are can be neglected in the first approximation.
Recalling 
%\com{that} 
that $\tilA \sim \cn(\theta; 1/\sqrt{2})$, we can write the EOM in the form
\begin{align}
	\ddot{a}_{A+}  - \cn^2\theta a_{A+} = 0, \\
	\ddot{a}_{A-}  + \cn^2\theta a_{A-} + \calO(\theta^{-5/2}a^3_\eta) = 0,
	\label{A-_late} \\
	\calL^2_\eta a^3_\eta + 2\cn^2\theta a^3_\eta - \sqrt{2}i\tiltheta^{1/2}p_\eta\tilA a_{A-} = 0,
\end{align}
for the $A$-sector and
\begin{align}
	\ddot{a}_{B+} + 3\cn^2\theta a_{B+} = 0, \\
	\ddot{a}_{B-} - \cn^2\theta a_{B-} = 0,
\end{align}
for the $B$-sector.
The EOMs of the $C$ and $C^*$-sector are given by
\begin{align}
	\ddot{a}^3_x + \cn^2\theta a^3_x + \calO(\theta^{-5/2}a^1_\eta) = 0 ,\\
	\calL^2_\eta a^1_\eta + \cn^2\theta a^1_\eta + i\tiltheta^{1/2}p_\eta\tilA a^3_x = 0,	
\end{align}
and
\begin{align}
	\ddot{a}^3_y + \cn^2\theta a^3_y + \calO(\theta^{-5/2}a^2_\eta) = 0 ,\\
	\calL^2_\eta a^2_\eta + \cn^2\theta a^2_\eta - i\tiltheta^{1/2}p_\eta\tilA a^3_y = 0,
\end{align}
respectively.
Thus, we get Lam\'{e}'s equations when $p_\eta\neq0, p_T=0$:
$a_{A+}$, $a_{B-}$ and $a_{3\eta}$ have exponential instability and their growth rates are determined by the Floquet theory independent of $p_\eta$.
It should be noted that $a_{A-}$ has only linear instability in the first approximation,
but the last term in Eq.~\eqref{A-_late} can be relevant at later times because $a^3_\eta$ grows exponentially.

On the other hand, 
the contribution from the background field is washed out for finite $p_T$ modes.
In the first approximation, 
the dominant contribution with respect to $\theta$ is of the order $\calO(\theta)$.
The EOMs of the $D$ and $E$-sector are given by
\begin{align}
	\ddot{a}^1_x - \cn^2\theta a^2_y = 0, \\
	\ddot{a}^2_x + 2\theta p^2_x/3 a^2_y + \calO(\theta^{1/2}) = 0, \\
	\ddot{a}^3_x + \calO(\theta^{1/2}) = 0,
\end{align}
and
\begin{align}
	\ddot{a}^1_y + 2\theta p^2_x/3 a^1_y + \calO(\theta^{1/2}) = 0, \\
	\ddot{a}^2_y + \calO(\theta^{1/2}) = 0, \\
	\ddot{a}^3_y + 2\theta p^2_x/3 a^3_y  + \calO(\theta^{1/2}) = 0,
\end{align}
respectively.
The EOM of the $F$-sector
\begin{align}
	\calL^2_\eta a^1_\eta + 2\theta p^2_x/3 a^1_\eta + \calO(\theta^{1/2}) = 0, \\
	\calL^2_\eta a^3_\eta + 2\theta p^2_x/3 a^3_\eta + \calO(\theta^{1/2}) = 0.
\end{align}
and the EOM of the $G$-sector
\begin{align}
	\calL^2_\eta a^2_\eta + 2\theta p^2_x/3 a^2_\eta  = 0.
\end{align}
have the same form.
In summary, we find that $a^2_x, a^1_y$ and $a^3_y$ 
obey Airy equation: 
\begin{align}
	\ddot{a} + \frac{2}{3}\theta p^2_x a = 0,
\end{align}
whose solution is given by $a = c_1 \Ai ((-2p_x^2/3)^{1/3}\theta) + ...$.
The longitudinal components 
%\sout{follow}\com{satisfy} 
satisfy Bessel equation:
\begin{align}
	\ddot{a}^a_\eta - \frac{3}{\theta}\dot{a}^a_\eta + \frac{2}{3}\theta p^2_x a^a_\eta  = 0,
\end{align}
whose solution is given by $a^a_\eta = c_1 \theta^2 J_{4/3}(p_x \tiltheta^{3/2}) + ...$.
Therefore, these modes do not show exponential instability in this stage.

\section{Linear instability}\label{App:Floquet}
Floquet's theorem states that the stability of fluctuations is governed by $\tr M$.
If $|\tr M|=2$,
there are not only stable solution but also linear divergent solution. 
Here, we show the origin of the linear divergence. 
Suppose that $\mu_1$ and $\mu_2$ are the eigenvalues of $M$,
the condition $|\tr M| = 2$ is realized if and only if $\mu_1=\mu_2=\pm1$.
The monodromy matrix cannot be 
%\sout{diagonalized form}\com{semisimple} but 
semisimple but 
%\com{only has} 
only has
Jordan normal form due to the degeneracy of the characteristic multiplier.
If $\tr M = 2$, without loss of generality,
we can take
\begin{align}
	M = 
	\begin{pmatrix}
		1 & 1 \\ 0 & 1
	\end{pmatrix}.
\end{align}
The logarithm of $M$ is now given by
\begin{align}
	\log M = \log \lbb 1 + 
	\begin{pmatrix}
		0 & 1 \\ 0 & 0
	\end{pmatrix}
	\rbb
	=
	\begin{pmatrix}
		0 & 1 \\ 0 & 0
	\end{pmatrix} .
\end{align}
Finally, we get
\begin{align}
	\Phi(t) = F(t) \exp 
	\begin{pmatrix}
		0 & t/T \\ 0 & 0
	\end{pmatrix}
	= F(t)
	\begin{pmatrix}
		1 & t/T \\ 0 & 1
	\end{pmatrix},
\end{align}
where $F(t)$ is a periodic function.
Actually, one solution is periodic in time and the another solution 
%\sout{linearly diverges}\com{is linearly divergent}.
is linearly divergent.
The case of $\tr M = -2$ can be discussed in a parallel way and then, we get anti-periodic and linearly divergent solutions.
In general, the degeneracy occurs at the boundaries of the instability bands.

%%%%%%%%%%%%
%%%%%%%%%%%%%%%%%%%%%%%%%%%%%%%%%%%%%%%%%%%%%%%%%%%%%%%%%%%%%%
\section{Effective growth rate}\label{App:growth}
In Sec~\ref{sec:Stability},
we have 
%\sout{discussed}\com{shown} 
shown that the growth rates of finite $p_\eta$ modes approach towards the upper bound which   is determined by the Floquet analysis in a nonexpanding geometry. 
More quantitative discussion can be performed as
% \sout{following way}\com{follows}.
follows.

Suppose that $\Phi(\theta)$ is 
%\sout{the}\com{a}
a Wronskian matrix of the given second order differential equation and
let us define
\begin{align}
	M(\theta_0,\theta) = \Phi(\theta_0)^{-1} \Phi(\theta).
\end{align}
$M(\theta_0,T+\theta_0)$ is equivalent to the monodromy matrix for the periodic-driven system with period $T$.
Thus the straightforward generalization of the Floquet exponents are given by the eigenvalues of the $\Gamma(\theta) = \log M /(\theta - \theta_0)$, say $\{\gamma_i\}$.
We call them 
%\sout{by} 
effective growth rates.

\begin{figure}[bth]
	\PSfig{8.0cm}{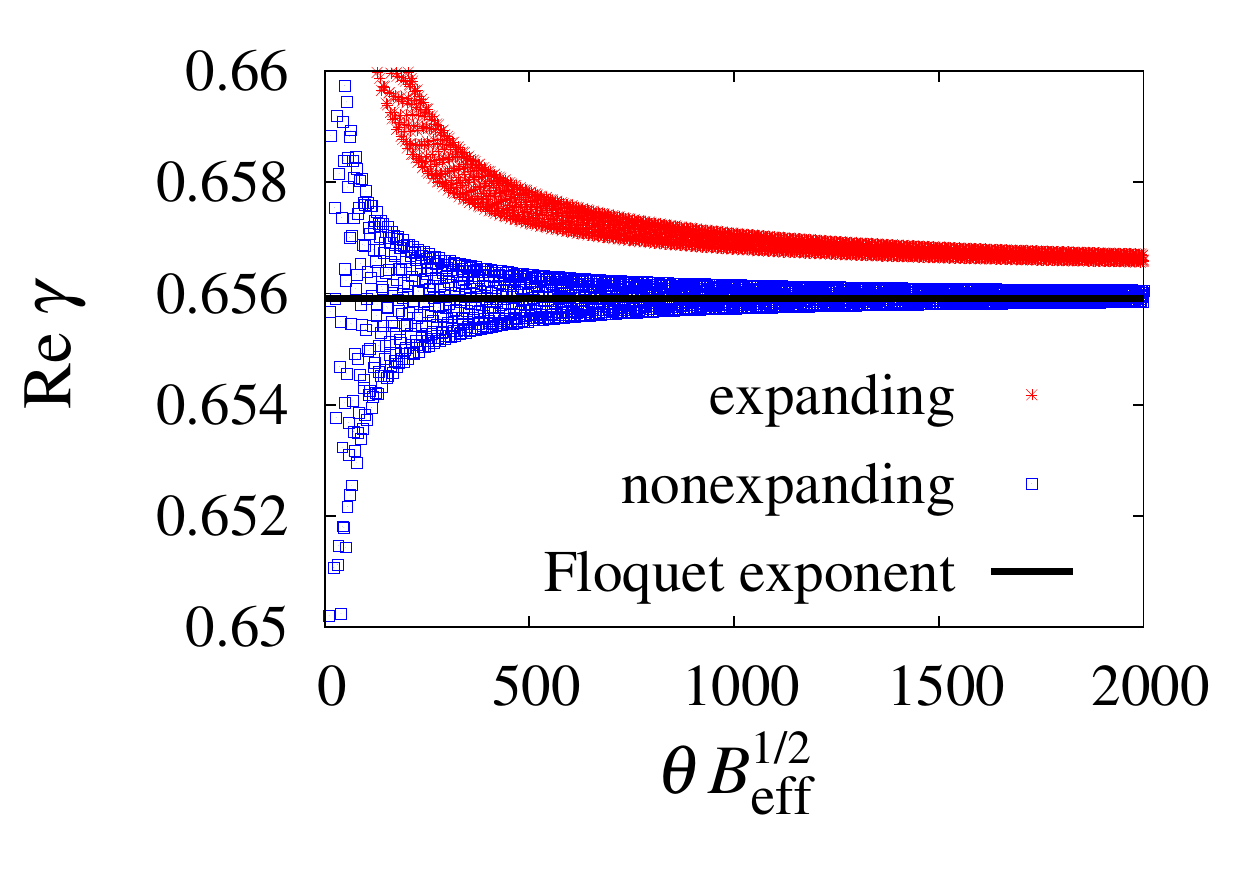}
	\caption{
		The effective growth rate of the $a_{A+}(p=0)$ (expanding) and Lam\'{e}'s equation with $\lambda=-1$ (nonexpanding).
		The black solid line describes the Floquet exponent.
	}
	\label{Fig:exponent}
\end{figure}

We calculate the maximal effective growth rate for $a_{A+}(p=0)$ and Lam\'{e}'s equation with $\lambda=-1$ which is the nonexpanding counterpart.
Figure~\ref{Fig:exponent} shows that the real part of the effective growth rate as a function of the conformal time.
The effective growth rate of the Lam\'{e}'s equation well agrees with the Floquet exponent whose value is approximately $0.6559$ as it should be.
The effective growth rate of $a_{A+}(p=0)$ is slightly larger than that of Lam\'{e}'s equation,
but it approaches toward the Floquet exponent.

\bibliographystyle{h-physrev5}
\bibliography{Expandref}
\end{document}